\documentclass[paper,notoc]{JHEP3}
\usepackage{epsfig}

\title{Precise evaluation of the Cut Constructible part of one loop amplitudes within D-dimensional unitarity.}
\author{Achilleas Lazopoulos\\
  Institute for Theoretical Physics, ETH Zurich,
  8093 Zurich, Switzerland\\
  E-mail: \email{lazopoulos@itp.phys.ethz.ch}
  }

\abstract{ 
A method to separate pentagon contributions from the evaluation of the cut constructible part of primitive amplitudes within the framework of D-dimensional unitarity is proposed. The cut constructible part is thus reconstructed with significantly higher accuracy as demonstrated by the numerical analysis presented here. }


\newcommand{\be}{\begin{equation}}
\newcommand{\ee}{\end{equation}}

\newcommand{\bea}{\begin{eqnarray}}
\newcommand{\eea}{\end{eqnarray}}

\newcommand{\nn}{\nonumber\\}

\begin{document}

\section{Introduction}
Precise predictions for signals and backgrounds at hadron colliders are the necessary ingredient for the unveiling of new physics. Such predictions can only be achieved with calculations beyond the tree-level leading order in the perturbation expansion over the strong coupling constant. 

Next to leading order QCD calculations consist of a virtual part and a real emission part. The real emission is nowadays automated either in the Catani-Seymour dipole approach \cite{Catani:1996vz}, implemented 
in \cite{Gleisberg:2007md},\cite{Seymour:2008mu},\cite{Hasegawa:2008ae,Hasegawa:2009tx},\cite{Frederix:2008hu},\cite{Czakon:2009ss}, 
or the Frixione-Kunszt-Signer approach\cite{Frixione:1995ms,Frixione:1997np} recently 
implemented in \cite{Frederix:2009yq}. The virtual part that involves one loop integrals has been tackled in many different ways, most of which involve some kind of reduction to master integrals. Each one loop integral is expressed as a linear combination of a small number of simpler integrals which are known in a closed analytic form. Traditionally this reduction is performed analytically for each one-loop Feynman diagram appearing in the calculation and the coefficients are known as functions of the kinematical variables of the process. There are many generic frameworks implemented that have been used the last years in very demanding calculations (see 
for example \cite{Denner:2006fy,Denner:2005nn,Denner:2002ii} 
used in \cite{Bredenstein:2008zb,Bredenstein:2008ia,Bredenstein:2009aj} 
or the publically available  \cite{Binoth:2008uq} used in \cite{Binoth:2009wk,Binoth:2009rv}).

Alternative approaches based on unitarity have been employed extensively in the past to recover one loop amplitudes (see e.g. \cite{Bern:1994cg,Bern:1994zx}). Following a generalization of the unitarity idea to quadruple cuts \cite{Britto:2004nc} and a novel approach in reduction that allows one to partially fraction numerically one loop integrals at the integrand level on a per point base \cite{Ossola:2006us, Ossola:2007ax}, the numerical evaluation of one loop amplitudes using tree-level quantities as building blocks appeared to be possible and efficient enough to tackle processes with many final state partons. Since then there have been mainly three approaches developed along the lines of numerical reduction with tree-level objects as building blocks: the Black Hat approach \cite{Berger:2006cz,Berger:2008sj}(based also on the reduction method of \cite{Forde:2007mi}, implemeneted in \cite{Berger:2008sj,Berger:2008ag}), the D-dimensional unitarity approach \cite{Ellis:2007br,Giele:2008ve,Ellis:2008kd,Ellis:2008ir} (implemented in various independent codes \cite{Giele:2008bc, Winter:2009kd, Melnikov:2009dn,Ellis:2008qc,Lazopoulos:2008ex} ) and the helac-1loop approach\cite{Ossola:2007ax,Ossola:2008xq,Ossola:2007zz,Ossola:2008zza,Ossola:2008zzb,Draggiotis:2009yb,Garzelli:2009is} (implemented in \cite{vanHameren:2009dr}). All three approaches have already delivered  differential cross sections for processes of prime phenomenological importance \cite{Berger:2008sz,Berger:2009zg,Berger:2009ba,Berger:2009ep,Berger:2009xp,Bevilacqua:2009zn,Ossola:2007bb, Melnikov:2009dn, KeithEllis:2009bu, Melnikov:2009wh, Ellis:2009zw}. 

Due to their infrared and ultraviolet singular nature, loop amplitudes need to be evaluated in $D=4-2\epsilon$ dimensions, and also the degrees of freedom of unobserved particles must be set according to  some regularization scheme.  Within the framework of D-dimensional unitarity, the virtual amplitude is considered as a function of the dimensionality of the virtual particles circulating in the loop.  Keeping the dimensionality integer, one can evaluate both the coefficients of the  cut constructible (CC) part and the full rational part of the one loop amplitude employing D-dimensional cuts that decompose the coefficients into tree-level amplitudes in D-dimensions. The price paid for the simplicity of the algorithm is the necessary inclusion of pentuple-cuts (absent in other approaches). 

Pentuple-cut residues involve the loop momentum in a high power and can pose a threat to the numerical accuracy of the result when evaluating the reduction coefficients for phase space points of degenerate kinematics (when any Gram determinant of the external momenta approaches zero). The percentage of such points varies with the process and the  kinematical cuts on the external partons employed to avoid soft and collinear singular regions. As a rule of thumb, their number increases in processes with many gluons and few fermion pairs. 

In particular, pentuple-cut residues when large, are seen to reduce the numerical accuracy of the rational as well as the cut-constructible part of the 
amplitude. On the other hand, it is known \cite{Ellis:2007br} that the CC part can be fully reconstructed by restricting the loop momentum (and therefore the whole unitarity method) in four dimensions. It is the aim of this work to show how the pentuple-cut residues can be  decoupled from the evaluation of the CC part which is then evaluated with much higher precision. 

\section{D-dimensional unitarity}

We will follow the version of D-dimensional unitarity presented in \cite{Ellis:2008kd}  and the notation of \cite{Lazopoulos:2008ex}. We employ color decomposition to write the virtual amplitude as a linear combination of gauge-invariant primitive amplitudes\cite{Bern:1994zx}. 

Each such primitve amplitude is a sum of one-loop integrals that need to be dimensionally regularized. Apart from the dimensionality of the integration, that is continued from $D=4$ to $D=4-2\epsilon$ there is also  the dimensionality, $D_s$ of the internal, unobservable, particles, which depends on the particular regularization scheme: it is set to $4$ in the Four Dimensional Helicity scheme (FDH)~\cite{Bern:2002zk} or to $4-2\epsilon$ in the 't Hooft - Veltman scheme (HV)~\cite{'tHooft:1972fi}. The dependence of the amplitude  on $D_s$ is linear (with the exception of fermion loops). Seen as a function of $D_s$, 
\be
A^{D_s}=A_0+A_1\cdot D_s
\ee
and one can 
 evaluate numerically  $A^{D_s}$ for two values of $D_s=D_{1,2}$, determine the $D_s$ independent $A_0,A_1$ from $A^{D_1}$ and $A^{D_2}$, and thus get the full $A^{D_s}$ for arbitrary $D_s$. Thereafter, setting $D_s$ to $4$ recovers the FDH scheme  and to $4-2\epsilon$ the 't Hooft - Veltman scheme. During all this the dimensionality of the loop momentum is kept to arbitrary $D$ with the constraint $D<D_s$. Only after the reduction is performed one can set $D\rightarrow 4-2\epsilon$ and evaluate the master integrals as a series in $\epsilon$. The problem of numerically evaluating full one loop amplitudes (including the rational part) reduces then to the problem of evaluating $A^{D_{1,2}}$. In order to accommodate fermions, one has to choose $D_{1,2}=6,8$. 

Each such primitive amplitude $A^{D_{1,2}}$ is partially fractioned in terms of a complete set of master integrals.
\bea
A^{D_{1,2}}(p_1,p_2,\ldots,p_N)=
 \int[dl] 
 & &
 \left\{ \sum_{J_5}{\frac{\bar{e}_{J_5}(l)}{\prod_{k\in J_5}D_{k}(l)}}
+\sum_{J_4}{ \frac{\bar{d}_{J_4}(l)}{\prod_{k\in J_4}D_{k}(l)}} \right.
\nn & &
\left.
+\sum_{J_3}{ \frac{\bar{c}_{J_3}(l)}{\prod_{k\in J_3}D_{k}(l)}}
+\sum_{J_2}{ \frac{\bar{b}_{J_2}(l)}{\prod_{k\in J_2}D_{k}(l)}}
\right\}
\eea
where $J_k=\{i_1,i_2,\ldots,i_k\}$ a set of $k$ indices corresponding to $k$ of the propagators that appear in the primitive amplitude ($J_k$ runs over the entire set of possible subsets of $k$ propagators among those that appear in the primitive)
. 
We parametrize the unintegrated coefficients $\bar{e}(l),\bar{d}(l),\bar{c}(l),\bar{b}(l)$ as follows
\bea
\bar{e}_{J_5}(l)&=&e_{J_5,0}\nn
\bar{d}_{J_4}(l)&=&
d_{J_4,0}+
d_{J_4,1}a_1+
d_{J_4,2}a_5^2+
d_{J_4,3}a_1a_5^2+
d_{J_4,4}a_5^4
\nn
\bar{c}_{J_3}(l)&=&
c_{J_3,0}+
c_{J_3,1} a_1+
c_{J_3,2} a_2 + 
c_{J_3,3}(a_1^2-a_2^2)+a_1a_2(
c_{J_3,4}+
c_{J_3,5}a_1+
c_{J_3,6}a_2) + 
\nn
& &
c_{J_3,7} a_1a_5^2+
c_{J_3,8}a_2a_5^2+
c_{J_3,9}a_5^4
\nn
\bar{b}_{J_2}(l)&=&
b_{J_2,0}+
b_{J_2,1}a_1+
b_{J_2,2}a_2+
b_{J_2,3}a_3+
b_{J_2,4}(a_1^2-a_3^2)+
b_{J_2,5}(a_2^2-a_3^2)+
b_{J_2,6}a_1a_2+
\nn
& &
b_{J_2,7}a_1a_3+
b_{J_2,8}a_2a_3+
b_{J_2,9}a_5^2
\nn
\label{parametrization}
\eea
with 
$
a_i=l\cdot n_i
$
and $n_i$ are the unit vectors that span the subspace that is orthogonal to the physical subspace spanned by the momenta involved in the propagators $J_k$. 

Once all coefficients are recovered the primitive amplitude can be written as 
\be
A=A_{cc}+A_{R}
\ee
\be
A_{cc}=\sum_{J_4}\tilde{d}_{J_4,0}I_{J_4}+\sum_{J_3}c_{J_3,0}I_{J_3}+\sum_{J_2} b_{J_2,0}I_{J_2}
\ee
where $\tilde{d}_{J_4,0}$ includes $d_{J_4,0}$ as well as contributions from pentagon scalar integrals that are reduced to boxes.
\be
A_R=-\sum_{J_4}\frac{d_{J_4,4}}{6}-\sum_{J_3}\frac{c_{J_3,9}}{2}-\sum_{J_2}\frac{b_{J_2,9}}{6}
\ee 

\section{System of equations and Pentagon contamination}

In order to evaluate the unknown coefficients one has to solve an OPP-like system: begining from the pentagon, for every $J_k$ a set of specific loop momenta $\hat{l}_{J_k;s}$ for which $D_k(\hat{l}_{J_k;s})=0$ for all $k\in J_k$ is determined. For those loop momenta, the corresponding propagators are then on-shell, and the pentuple-cut coefficient $\bar{e}(l_s)$ can be determined as the residue of the loop integral, which is seen to be equal to a product of tree-level color-ordered amplitudes. Given the pentagon coefficient, one can proceed to the boxes. The box coefficients $\bar{d}(l)$ is equal to the residue minus a counter term from pentagons. Then one proceeds to the triple cuts and so on and so forth. This is what is meant by the statement that the system of OPP equations is triangular (see \cite{Ossola:2006us, Ossola:2007ax}). 

The system of equations we have to solve is, therefore,

\bea
\bar{e}_{I_5,s} & = & \left. \mathbb{R}\right|^{I_5}_{l_s}  =  e_{I_5,0} \nn
\bar{d}_{I_4,s}      & = & \mathbb{R}|^{I_4}_{l_s}	 
				-\sum_{J_{5|I_4}} \frac{\bar{e}_{J_5 }}{D_{J_5/I_4}(l_s)}  
					=  \sum_{r=0\dots 4} d_{I_4,r}f_r(l_s)\nn
\bar{c}_{I_3,s}      & = & \mathbb{R}|^{I_3}_{l_s} 	
				-\sum_{J_{5 | I_3}} \frac{\bar{e}_{J_5 }}{\prod_{k \in J_5/I_3}D_{k}(l_s) }
				-\sum_{{J_4 | I_3} }\frac{\bar{d}_{J_4 }(l_s)}{D_{J_4/I_3}(l_s)}  
					=  \sum_{r=0\ldots 9} c_{I_3,r}g_r(l_s) 		\nn
\bar{b}_{I_2,s}      & = & \mathbb{R}|^{I_2}_{l_s} 
				-\sum_{J_{5 | I_2}} \frac{\bar{e}_{J_5 }}{\prod_{k\in J_5/I_2}D_{k}(l_s) }
				-\sum_{J_{4 | I_2}}\frac{\bar{d}_{J_4}(l_s)}{\prod_{k\in J_4/I_2}D_{k}(l_s)} 
				-\sum_{J_{3 | I_2}}\frac{\bar{c}_{J_3 }(l_s)}{D_{J_3/I_2}(l_s)}  
					=  \sum_{r=0\ldots 9} b_{I_2,r}h_r(l_s) \nn
\label{system}
\eea
where $\mathbb{R}|^{I_k}_{l_s}$ is the value of the residue of the loop amplitude with respect to the propagators indicated by $I_k$ and the choice of loop momentum $l_s$. Moreover, 
\be
J_{a | I_b} \equiv J_a : J_a \cap I_b = I_b
\ee
The right hand side of the above equations is just a shorthand for the parametrization
of eq.\ref{parametrization}, i.e. the functions $f_r(l), g_r(l),h_r(l)$ are just the quantities multiplying the coefficients $d_{I_4,r},c_{I_3,r},b_{I_2,r}$ in eq.\ref{parametrization}.

Among the unknown coefficients $\{e_0,d_r,c_r,b_r\}$ the subset $\{ d_{0,1},c_{0\ldots6},b_{0\ldots8} \}$ can be evaluated separately, by choosing the loop momentum $l_s$ in four dimensions. In that case  $\{d_{3,4,5},c_{7,8,9},b_9\}$ drop out. The subset $\{ d_{0,1},c_{0\ldots6},b_{0\ldots8} \}$  reconstruct the four-dimensional part of the amplitude and $d_0,c_0,b_0$ are the coefficients of the master integrals in the cut constructible part of the amplitude. Therefore their value should be independent of the value of pentuple cuts. This however is not explicit in the above system, where pentuple coefficients are involved as counterterms in the equations for quadruple-, triple- and double-cut coefficients. 

That this is an undesirable situation is made clear from studies of numerically unstable points.  Pentuple-cut coefficients are carrying the effect of possible small Gram determinants to the fifth power. In the framework of the Vermaseren- van Neerven base ~\cite{Ellis:2007br},\cite{vanNeerven:1983vr}, Gram determinants appear as denominators in the evaluation of the loop momentum part that is fixed to solve the unitarity constraints of the cut. Phase space points for which the Gram determinants are approaching zero lead to numerically large components of the pentuple-cut loop momentum and therefore to large pentagon coefficients.  Their effect must cancel, however, when solving the four-dimensional subsystem of eq.~\ref{system} since the cut-constructible master integral coefficients can be derived by the purely four-dimensional of ~\cite{Ellis:2007br} where there are no pentagons. 

The cancellation of the pentagon contribution within the cut-constructible sub-system is the main source of numerical instabilities in the calculation of the CC part of the amplitude. One can see the correlation of numerically unstable phase space points (as detected by the inaccurate reconstruction of their pole coefficients in the $\epsilon=2-D/2$ expansion of the amplitude) with the appearance of small Gram determinants in the pentuple cut or equivalently with the appearance of large pentagon coefficients. This has already been noted in \cite{Winter:2009kd}. 

\EPSFIGURE{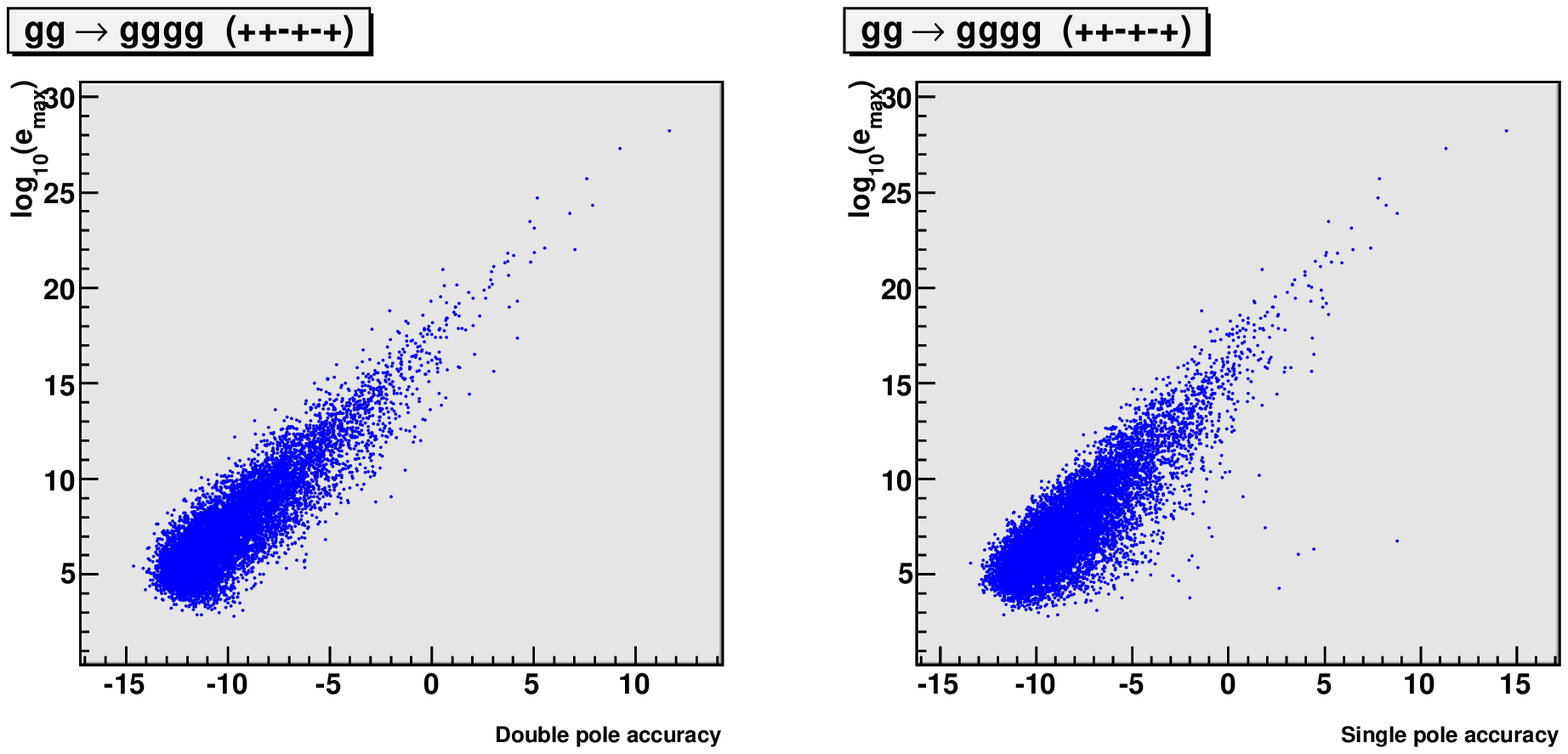,width=0.95\textwidth}
{
Correlation of the maximum pentagon coefficient with the double and single pole accuracies defined in the text 
   $\epsilon_{2,1}=\log_{10} \left|
   				\frac{  E_{1,2}-E^{an}_{1,2}  }  {  E^{an}_{1,2} } 
   				\right|$. 
			See section \ref{Kinematical_cuts} for details about the generation of the phase space pointset. 
\label{fig:scatter_old}
}


It is also shown in the plots of fig.\ref{fig:scatter_old} for the case of six gluons. Very similar correlation patterns are found for more external particles and for amplitudes with external fermions.

\DOUBLEFIGURE
{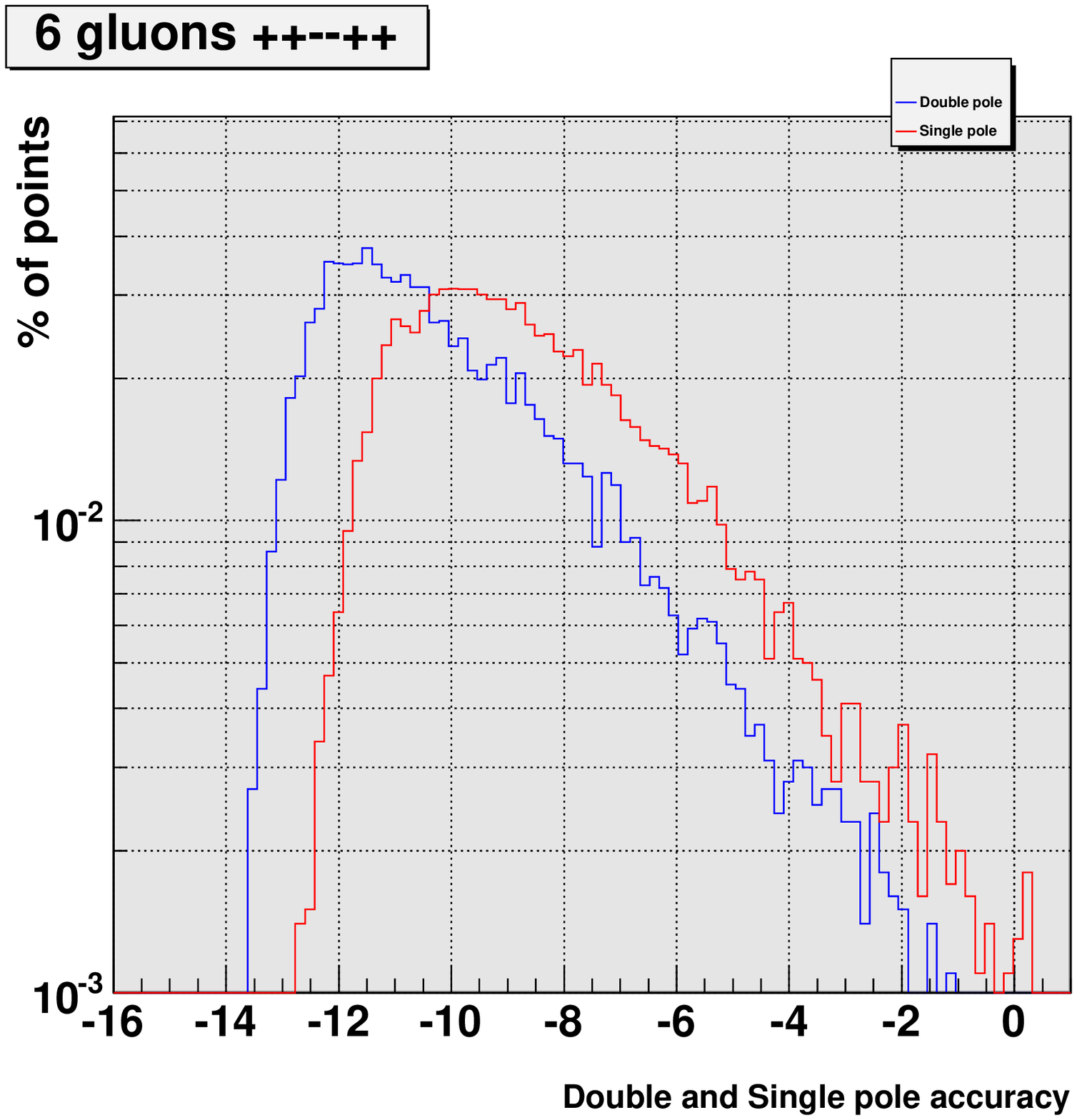,width=0.47\textwidth}
{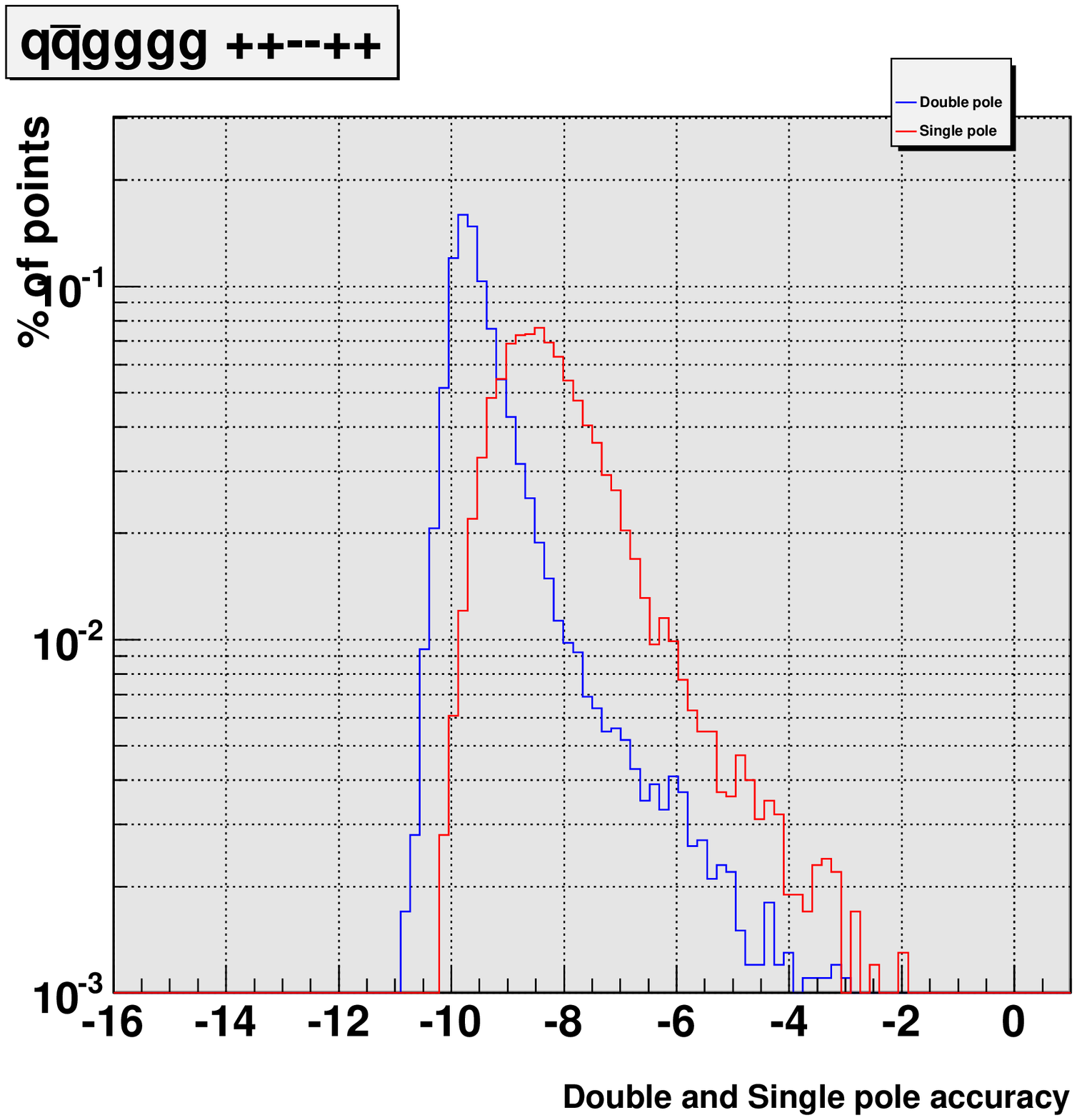,width=0.47\textwidth}
{Double and Single pole accuracy in $N=6$ gluons with the canonical approach.See section \ref{Kinematical_cuts} for details about the generation of the phase space pointset.\label{6g_poles_old}
}
{Double and Single pole accuracy in $q\bar{q}\rightarrow 4g$ with the canonical approach.See section \ref{Kinematical_cuts} for details about the generation of the phase space pointset.\label{qq4g_poles_old}
}


\DOUBLEFIGURE
{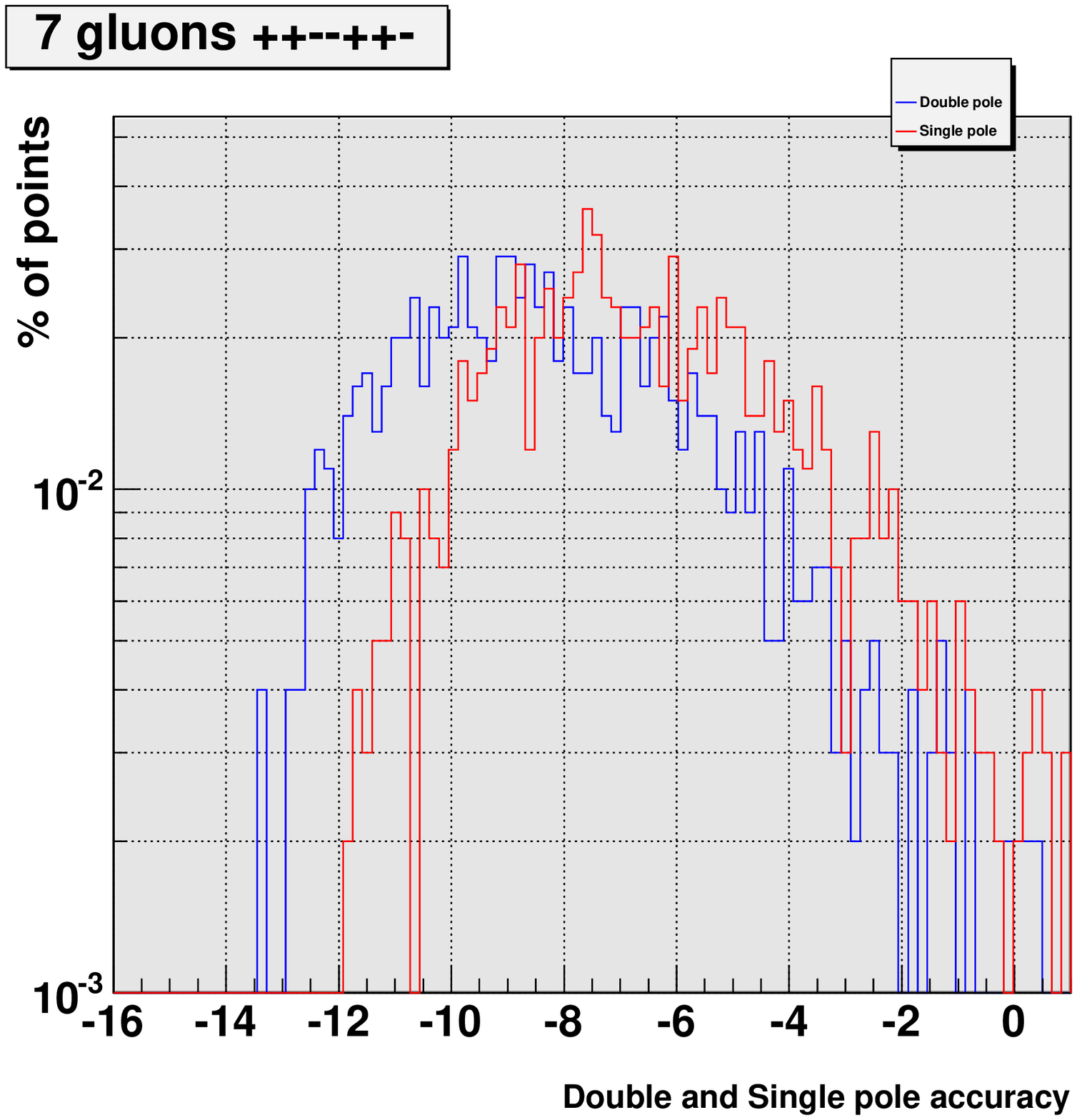,width=0.47\textwidth}
{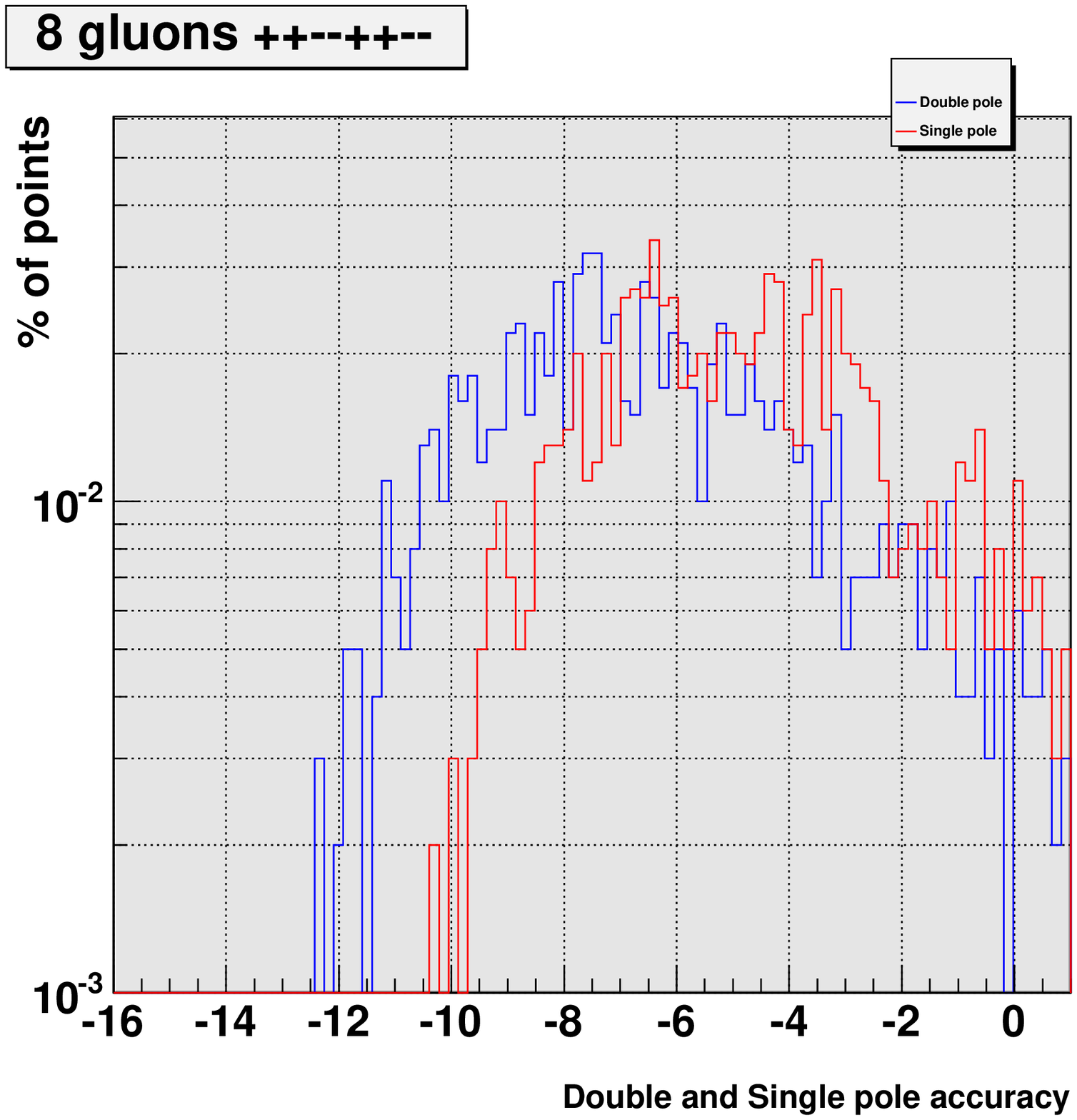,width=0.47\textwidth}
{
Double and Single pole in $N=7$ gluons with the canonical approach.
See section \ref{Kinematical_cuts} for details about the generation of the phase space pointset.
\label{7g_poles_old}
}
{Double and Single pole in $N=8$ gluons with the canonical approach.
See section \ref{Kinematical_cuts} for details about the generation of the phase space pointset.
\label{8g_poles_old}
}


The effect of pentagon contamination in the calculation of the CC part of the amplitude is exhibited in fig.\ref{6g_poles_old}-\ref{qq4g_poles_old} for the six gluon primitive and the $q\bar{q}\to 4g$ leading order primitive. The percentage of points that can be classified as numerically unstable based on the pole accuracy criterion varies from $5\%$ downwards, assuming reasonable final state cuts (e.g. a cone separation angle of $R=0.7$ and a $p_T$ cut of $20$GeV). If one would like to explore the collinear or soft limits, of course, the number of unstable points would increase.

\section{Pentagon isolation}
One can adopt the attitude of completely discarding the problematic points (assuming that they don't populate particularly contingent regions of the phase space) or resorting to quadruple precision (with 32 or 64 digits in the floating point arithmetics) to evaluate them accurately. We would like to suggest in this paper a method to reduce the number of unstable points drastically.

Given that the system ~\ref{system} is linear, we split it in two systems of equations, isolating the counterterms that include pentagon coefficients.
\bea
\bar{d}_{I_4,s} &=& \bar{d}_{I_4,s}^{a}+\bar{d}_{I_4,s}^{b}  
\nn
			& &\bar{d}_{I_4,s}^{a}=\mathbb{R}|^{I_4}_{l_s} = \sum_r d_{I_4,r}^{a}f_r(l_s)
 \nn
			& & \bar{d}_{I_4,s}^{b}=-\sum_{J_{5 |  I_4}} \frac{\bar{e}_{J_5 }}{D_{J_5/I_4}(l_s)} = \sum_r d_{I_4,r}^{b}f_r(l_s)
\nn
\eea
\bea
\bar{c}_{I_3,s} 	&=&  \bar{c}_{I_3,s}^{a} + \bar{c}_{I_3,s}^{b}  
\nn
			& &	\bar{c}_{I_3,s}^{a}= \mathbb{R}|^{I_3}_{l_s} 
				-\sum_{J_{4 | I_3}}\frac{\bar{d}^{a}_{J_4}(l_s)}{D_{J_4/I_3}(l_s)} =  \sum_r c_{I_3,r}^{a}g_r(l_s) 
\nn
			& &	\bar{c}_{I_3,s}^{b}=-\sum_{J_{5  |  I_3}} \frac{\bar{e}_{J_5 }}{\prod_{k \in J_5/I_3}D_{k}(l_s)}  
								-\sum_{J_{4 | I_3}}\frac{\bar{d}^{b}_{J_4 / I_3}(l_s)}{D_{k1}(l_s)} 
								= \sum_r c_{I_3,r}^{b}g_r(l_s) \nn
\eea
\bea
\bar{b}_{I_2,s} &=& \bar{b}_{I_2,s}^{a} + \bar{b}_{I_2,s}^{b}  
\nn
			& & \bar{b}_{I_2,s}^{a} = \mathbb{R}|^{I_2}_{l_s} 
				-\sum_{J_{4|I_2}}\frac{\bar{d}^{a}_{J_4 }(l_s)}{\prod_{k \in J_4/I_2}D_{k}(l_s)} 
				-\sum_{J_{3|I_2}}\frac{\bar{c}^{a}_{J_3 }(l_s)}{D_{J_3/I_2}(l_s)} 
				=  \sum_r b_{I_2,r}^{a}h_r(l_s)  \nn
			& &	\bar{b}_{I_2,s}^{b} = -\sum_{J_{5 | I_2}} \frac{\bar{e}_{J_5 }}{\prod_{k \in J_5/I_2}D_{k}(l_s)} 
 								-\sum_{J_{4| I_2}}\frac{\bar{d}^{b}_{J_4 }(l_s)}{\prod_{k \in J_4/I_2}D_{k}(l_s)} 
								-\sum_{J_{3| I_2}}\frac{\bar{c}^{b}_{J_3 }(l_s)}{D_{J_3/I_2}(l_s)}
								 = \sum_r b_{I_2,r}^{b}h_r(l_s)\nn
\eea
The b-subsystem contains exclusively contributions coming from the pentagon either directly or through box and triangle counter terms, while the a-subsystem has contributions from quadruple-, triple- and double-cuts as well as counter terms coming from them.

When the loop momenta are chosen in 4-d the b-subsystem coefficients of the boxes $D^b_{0,1}$ are non-zero. In that case there is no freedom in the choice of the box loop momenta. There are two solutions of the box unitarity constraints, $l^{\mu}_{I_4;\pm}=V_{I_4}\pm \sqrt{-V_{I_4}^2}n^{\mu}_{I_4}$. The coefficient $d^b_{I_4,0}$ is then 
\be
d^b_{I_4,0}=\frac{1}{2}(\bar{d}_{I_4}^{b}(l_+)+\bar{d}_{I_4}^{b}(l_{-}))= 
-\frac{1}{2}\sum_{J_{5 |  I_4}} \bar{e}_{J_5 }\left(\frac{1}{D_{J_5/I_4}(l_+)}+\frac{1}{D_{J_5/I_4}(l_-)} \right)
\ee
However all pentagon master integrals are reduced to boxes, so the final coefficient of each box  master integral including contributions from the reduction of the pentagons is 
\be
d_{I_4,0;full}=d_{I_4,0}^a+d_{I_4,0}^b+\tilde{d}_{I_4,0}
\ee  
and it is easy to see that 
\be
\tilde{d}_{I_4,0}=-d_{I_4,0}^b=\frac{1}{2}\sum_{J_{5 |  I_4}} \bar{e}_{J_5 }\left(\frac{1}{D_{J_5/I_4}(l_+)}+\frac{1}{D_{J_5/I_4}(l_-)} \right)
\ee
This means that the box coefficient $d_{I_4,0,full}$ can be evaluated without the contribution of the pentagons when the loop momenta are in 4-d. 
Furthermore, we have 
\be
d^b_{I_4,1}=\frac{1}{2\sqrt{-V_{I_4}^2}}(\bar{d}_{I_4}^{b}(l_+)-\bar{d}_{I_4}^{b}(l_{-}))= 
-\frac{1}{2\sqrt{-V_{I_4}^2}}\sum_{J_{5 |  I_4}} \bar{e}_{J_5 }\left(\frac{1}{D_{J_5/I_4}(l_+)}-\frac{1}{D_{J_5/I_4}(l_-)} \right)
\ee
from which we get 
\be
\bar{d}_{I_4}(l_{\in 4D})=-\sum_{J_{5| I_4}} \bar{e}_{J_5}\frac{1}{2} \left(\frac{1+z_{I_4}(l)}{D_{J_5/I_4}(l_+)}+\frac{1-z_{I_4}(l)}{D_{J_5/I_4}(l_-)}\right)
\ee
with
\be
z_{J_5}(l)=\frac{l\cdot n_{I_4}}{\sqrt{-V_{I_4}^2}}
\ee
The b-subsystem for the triple coefficients is then 

\bea
\bar{c}_{I_3,s}^{b}&=&-\sum_{J_{5  |  I_3}} \frac{\bar{e}_{J_5 }}{\prod_{k \in J_5/I_3}D_{k}(l_s)}  
								-\sum_{J_{4 | I_3}}\frac{\bar{d}^{b}_{J_4 / I_3}(l_s)}{D_{k1}(l_s)} \nn
			  &=&-\sum_{J_{5  |  I_3}} \frac{\bar{e}_{J_5 }}{\prod_{k \in J_5/I_3}D_{k}(l_s)}
							+\sum_{J_{4 | I_3}}\frac{1}{D_{ J_4/I_3}(l_s)} \sum_{J_{5| J_4}} \bar{e}_{J_5}\frac{1}{2} 
								\left( \frac{1+z_{J_4}(l_s)}{D_{J_5/J_4}(l_+)}+\frac{1-z_{J_4}(l_s)}{D_{J_5/J_4}(l_-)} \right)
								\nn								
\eea
The summation symbols can be interchanged since $\sum_{J_{4|I_3}}\sum_{J_{5|I_4}}=\sum_{J_{4|I_3}}\sum_{J_{5|I_3}}$, so we have
\bea
\bar{c}_{I_3,s}^{b}&=&-\sum_{J_{5  |  I_3}} \frac{\bar{e}_{J_5 }}{\prod_{k \in J_5/I_3}D_{k}(l_s)}
							+\sum_{J_{5| I_3}} \bar{e}_{J_5}\sum_{J_{4 | I_3}}\frac{1}{D_{ J_4/I_3}(l_s)} \frac{1}{2} 
								\left( \frac{1+z_{J_4}(l_s)}{D_{J_5/J_4}(l_+)}+\frac{1-z_{J_4}(l_s)}{D_{J_5/J_4}(l_-)} \right)
								\nn	
			& = & -\sum_{J_{5  |  I_3}} \bar{e}_{J_5 }	\left[
											\frac{1}{\prod_{k \in J_5/I_3}D_{k}(l_s)}
												-\sum_{J_{4 | I_3}}\frac{1}{D_{ J_4/I_3}(l_s)} \frac{1}{2} 
								\left( \frac{1+z_{J_4}(l_s)}{D_{J_5/J_4}(l_+)}+\frac{1-z_{J_4}(l_s)}{D_{J_5/J_4}(l_-)} \right) \right] \nn
\eea
The quantity inside the brackets vanishes identically when $l_s$ is in 4D and it solves the unitarity constraints of the $I_3$ triple cut. Note that this cancellation is a kinematical identity and is therefore independent of whether the external or internal particles are fermions or gluons. The cancellation has been numerically checked to hold up to 12 digits in the double precision version of the code for processes with up to 8 external particles of all flavors.  

This means that all $\bar{c}^b_{I_3,s}$ vanish identically if $l_s$ is in 4D, and so do the triangle coefficients $c^b_{I_3,0\ldots 6}$ for all $I_3$. Similarly, the bubble coefficients $b^b_{I_2,0\ldots8}$ vanish identically.

Therefore the whole b-subsystem doesn't contribute to the CC part of the amplitude. It is, however, necessary for the evaluation of the rational part.

\DOUBLEFIGURE
{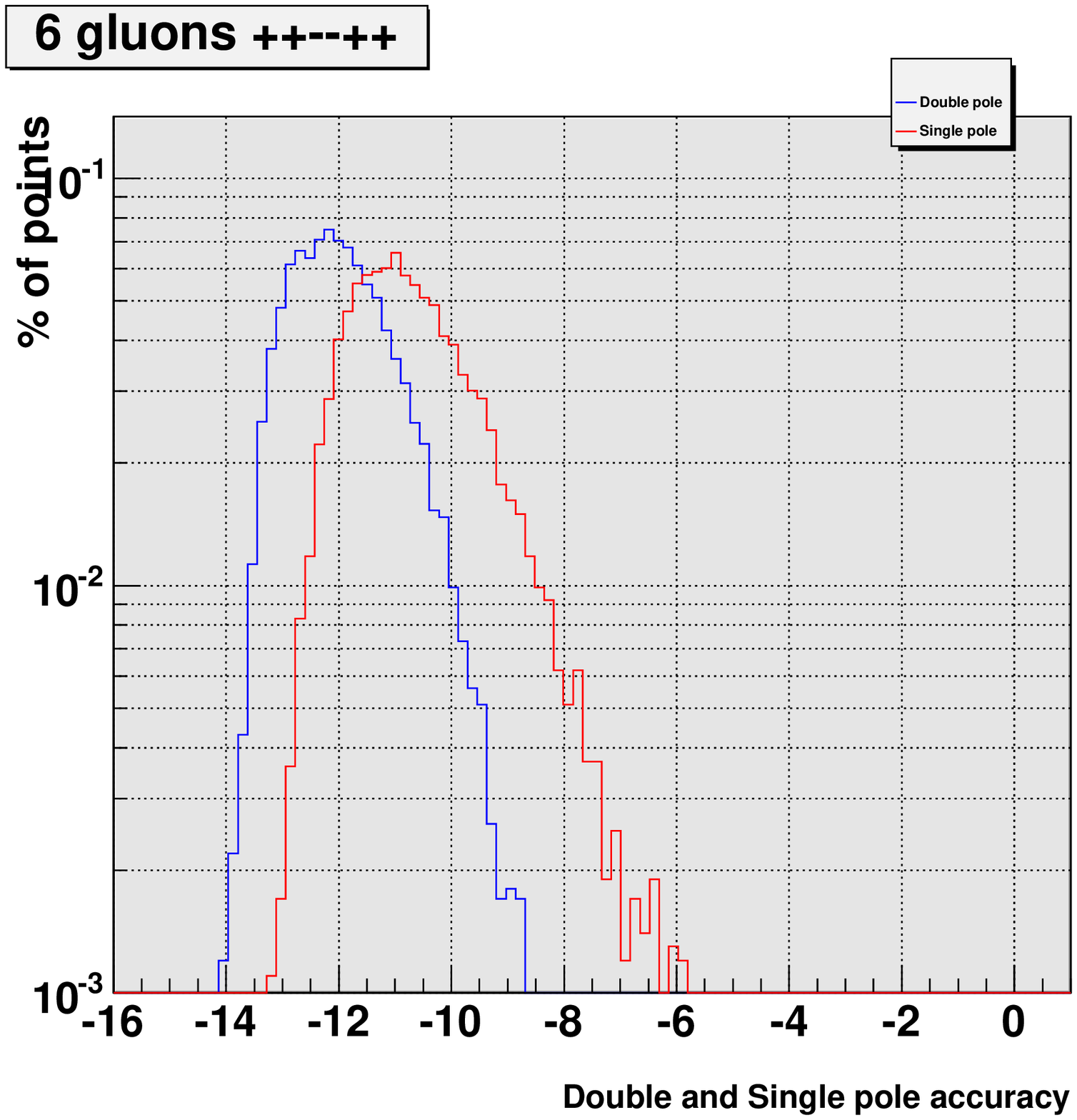,width=0.47\textwidth}
{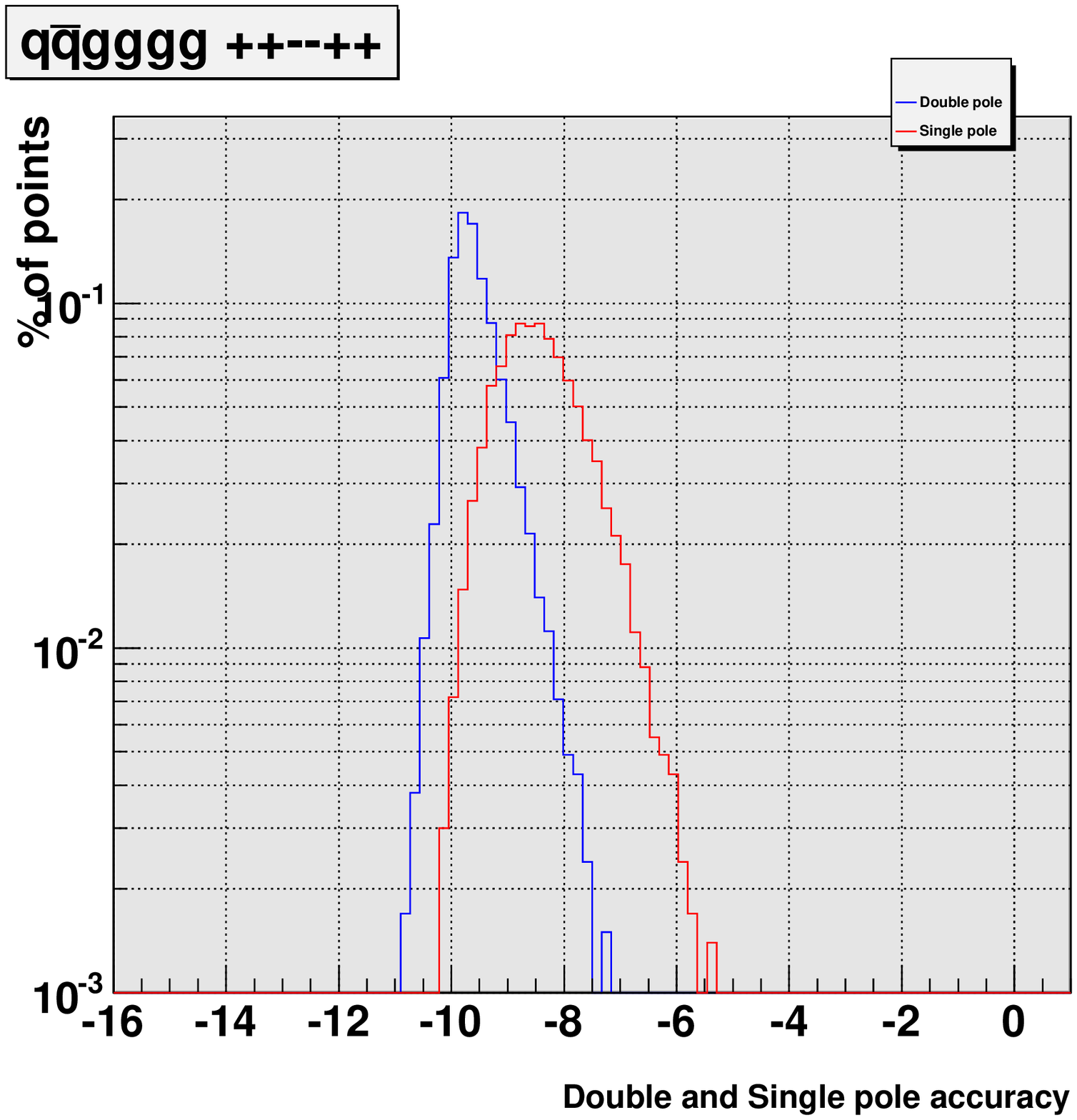,width=0.47\textwidth}
{
Double and Single pole in $N=6$ gluons with the \emph{new} approach.
See section \ref{Kinematical_cuts} for details about the generation of the phase space pointset.
\label{6g_poles_new}
}
{
Double and Single pole in $q\bar{q}\rightarrow 4g$ with the \emph{new} approach.See section \ref{Kinematical_cuts} for details about the generation of the phase space pointset.
\label{qq4g_poles_new}
}


\DOUBLEFIGURE
{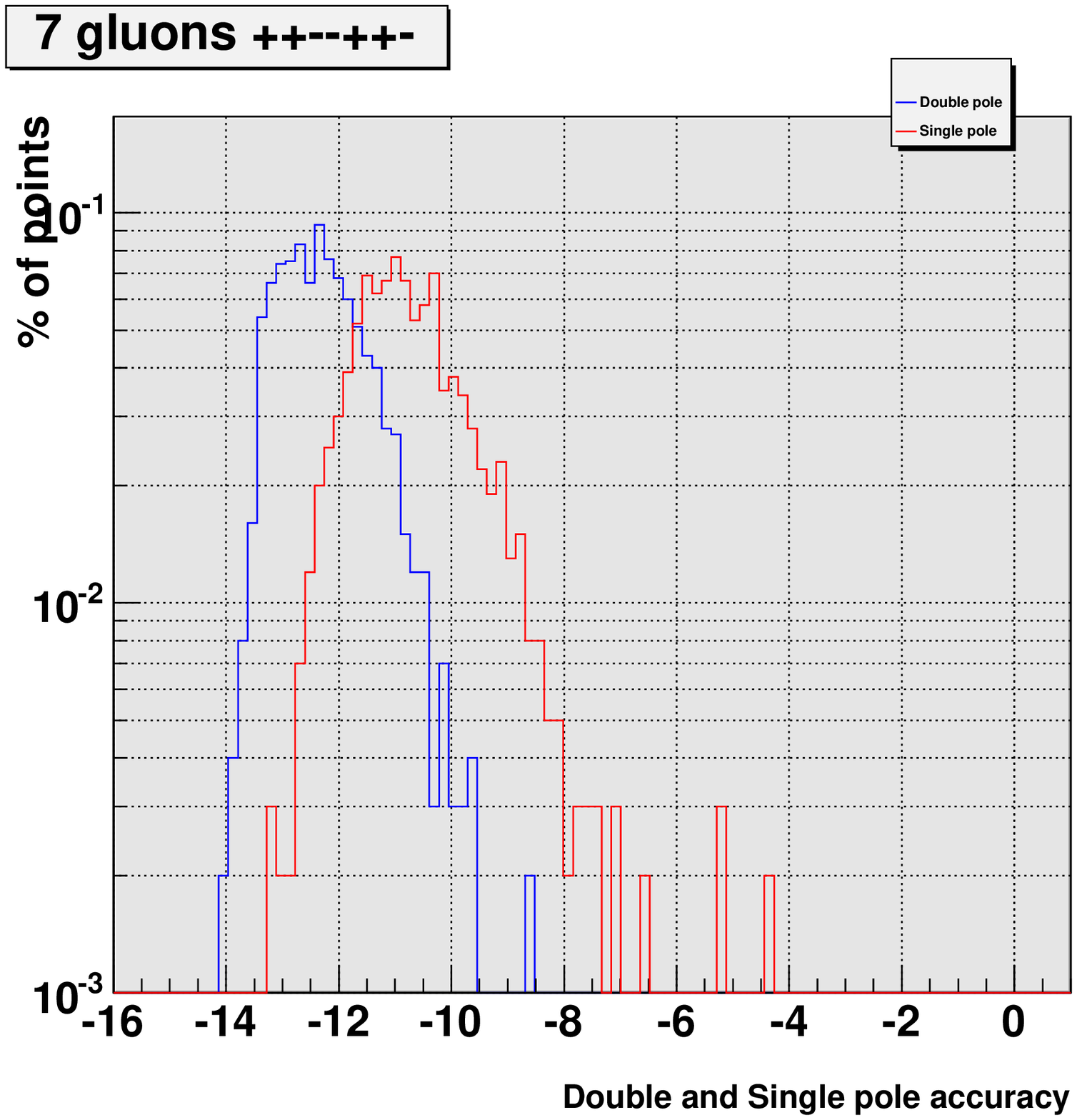,width=0.47\textwidth}
{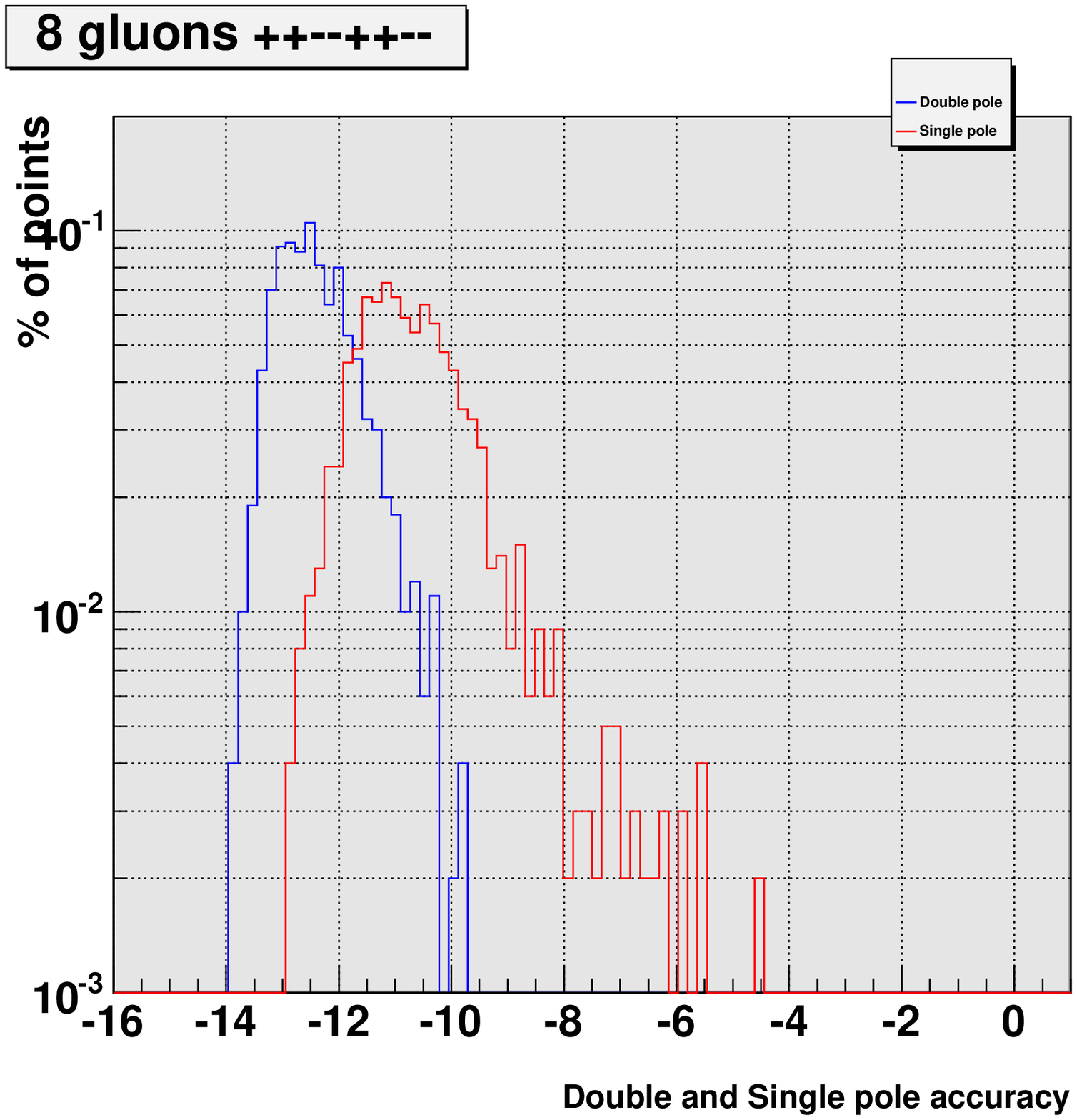,width=0.47\textwidth}
{
Double and Single pole in $N=7$ gluons with the \emph{new} approach.See section \ref{Kinematical_cuts} for details about the generation of the phase space pointset.
\label{7g_poles_new}
}
{
Double and Single pole in $N=8$ gluons with the \emph{new} approach.See section \ref{Kinematical_cuts} for details about the generation of the phase space pointset.
\label{8g_poles_new}
}

%

In this setup, one can thus disentangle completely the pentagon coefficients from the cut constructible part of the primitive amplitude. The difference this makes  in the numerical stability of the cut constructible part is rather spectacular as can be seen in figs.\ref{6g_poles_new},\ref{qq4g_poles_new},\ref{7g_poles_new},\ref{8g_poles_new}, where the double and single pole accuracy plots of fig.\ref{6g_poles_old},\ref{qq4g_poles_old},\ref{7g_poles_old},\ref{8g_poles_old} are revisited. The CC part is now evaluated by solving the two subsystems separately.  In figs \ref{cc_PID_1}-\ref{cc_PID_4} we show the accuracy of the CC part of the finite result for a primitive with gluons and the leading color primitive with one fermion pair, as compared with the quadruple precision result. 

The improvement is significant and  the percentage of points when one needs to go to higher precision in order to reconstruct the CC part are in the worst of cases at the per mille level.

\DOUBLEFIGURE
{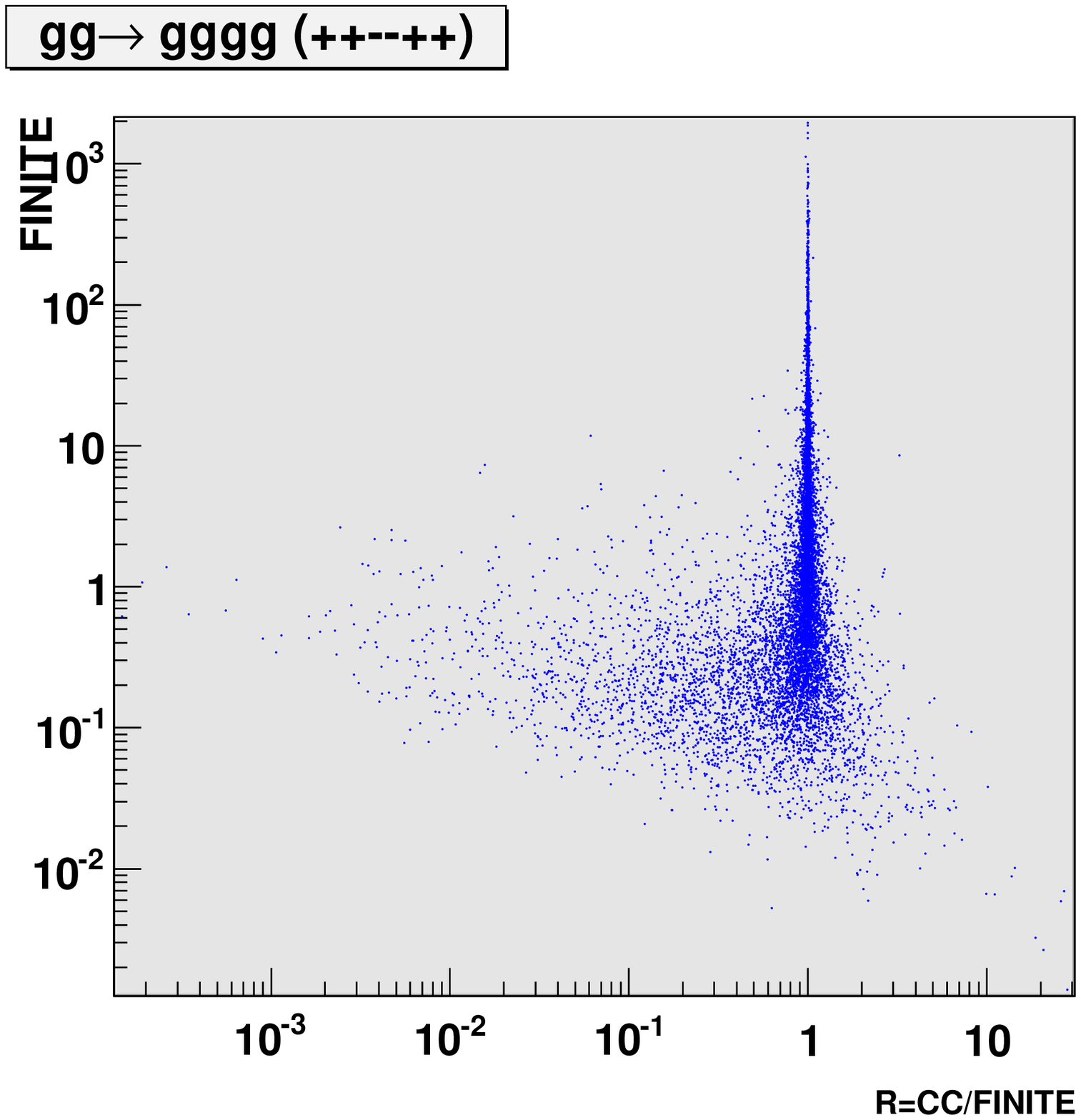,width=0.47\textwidth}
{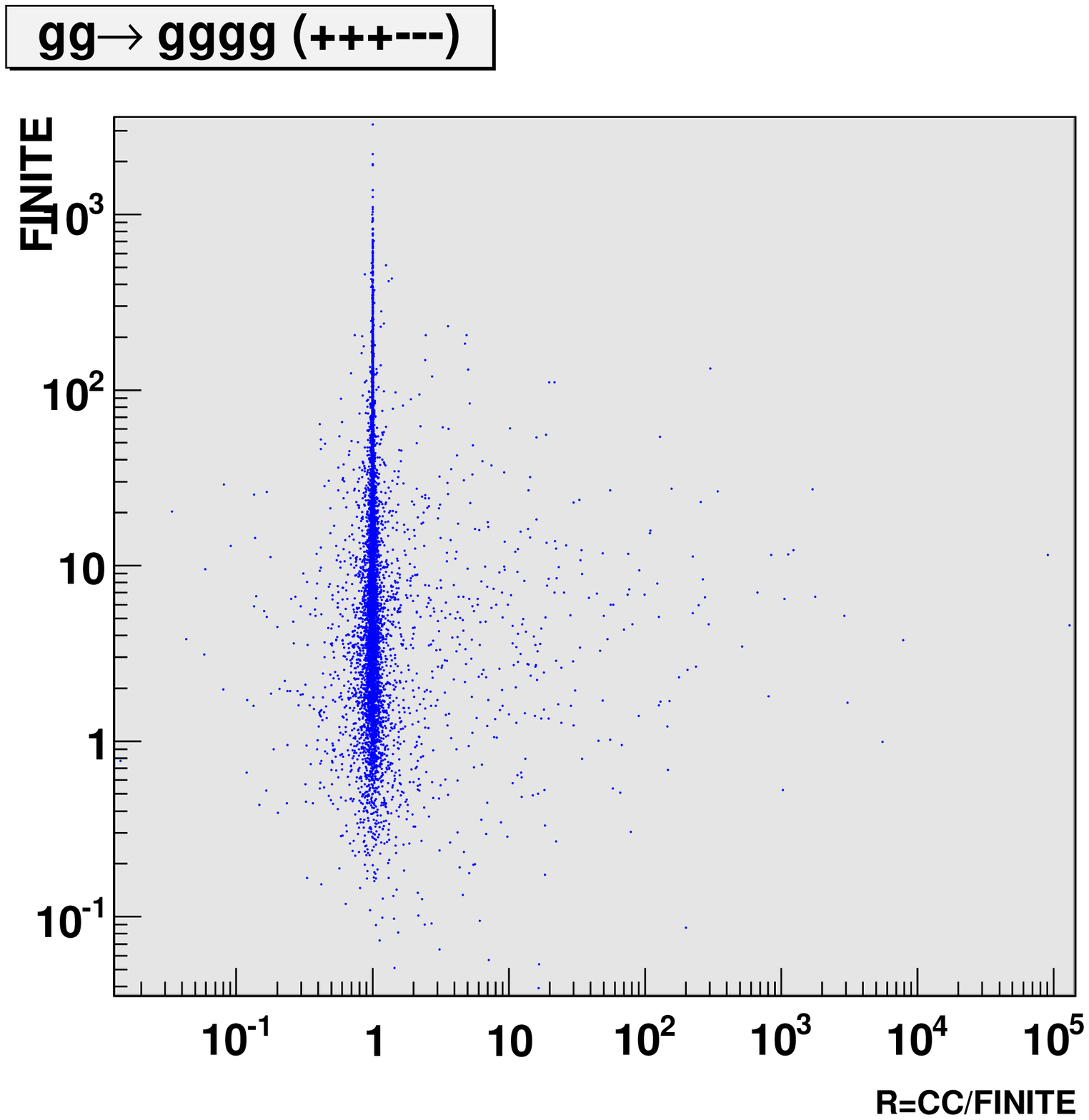,width=0.47\textwidth}
{
Scatterplot of the absolute value of the total finite result for the primitive amplitude, $|FIN|$ versus the ratio $|CC| /  |FIN|$ for the $++--++$ helicity configuration of the 6 gluon gluonic primitive. See section \ref{Kinematical_cuts} for details about the generation of the phase space pointset.
\label{cc_over_finite_1}
}
{
Scatterplot of the absolute value of the total finite result for the primitive amplitude, $|FIN|$ versus the ratio $|CC| /  |FIN|$ for the $+++---$ helicity configuration of the 6 gluon gluonic primitive. See section \ref{Kinematical_cuts} for details about the generation of the phase space pointset.
\label{cc_over_finite_2}
}


\DOUBLEFIGURE
{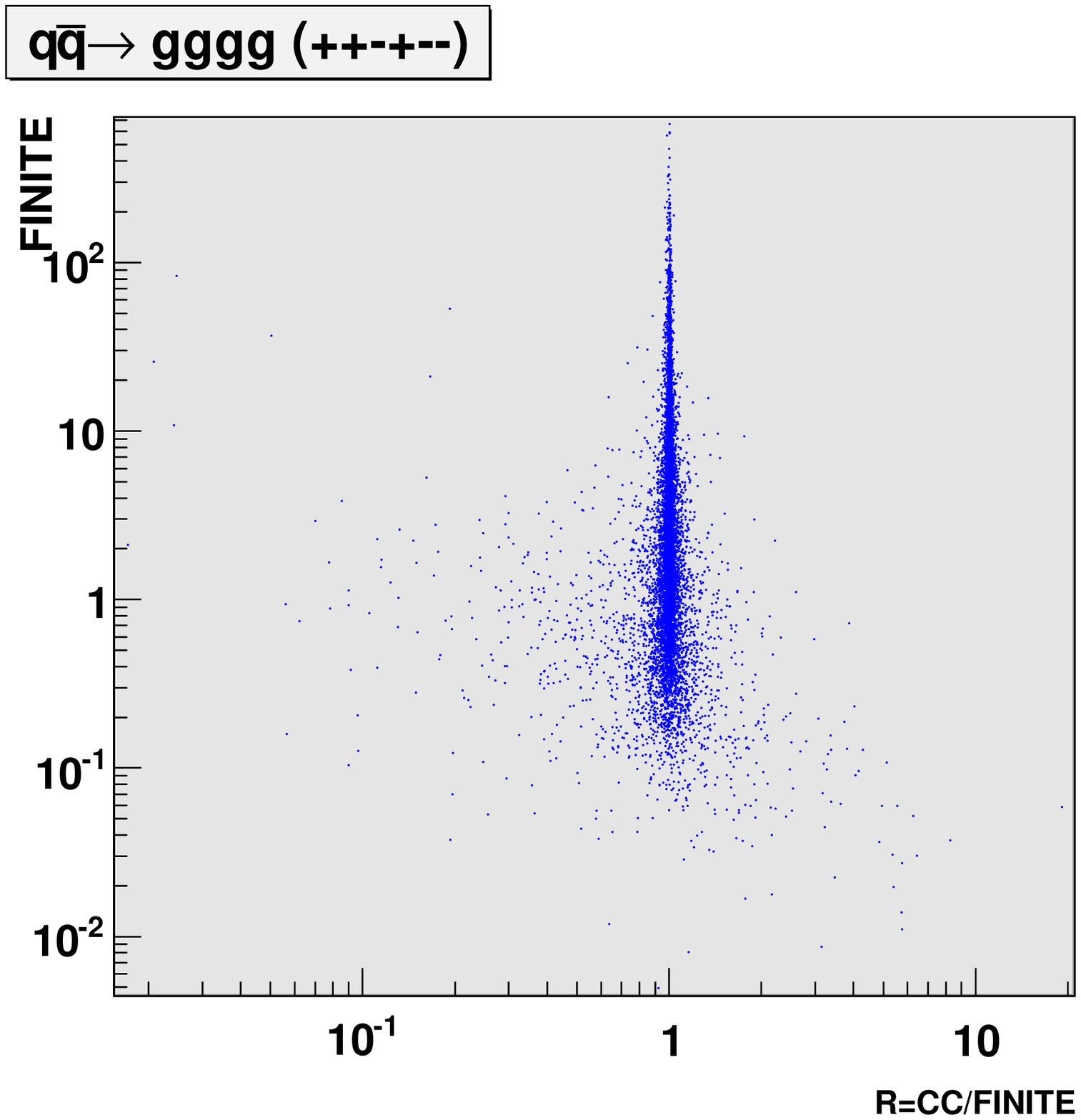,width=0.47\textwidth}
{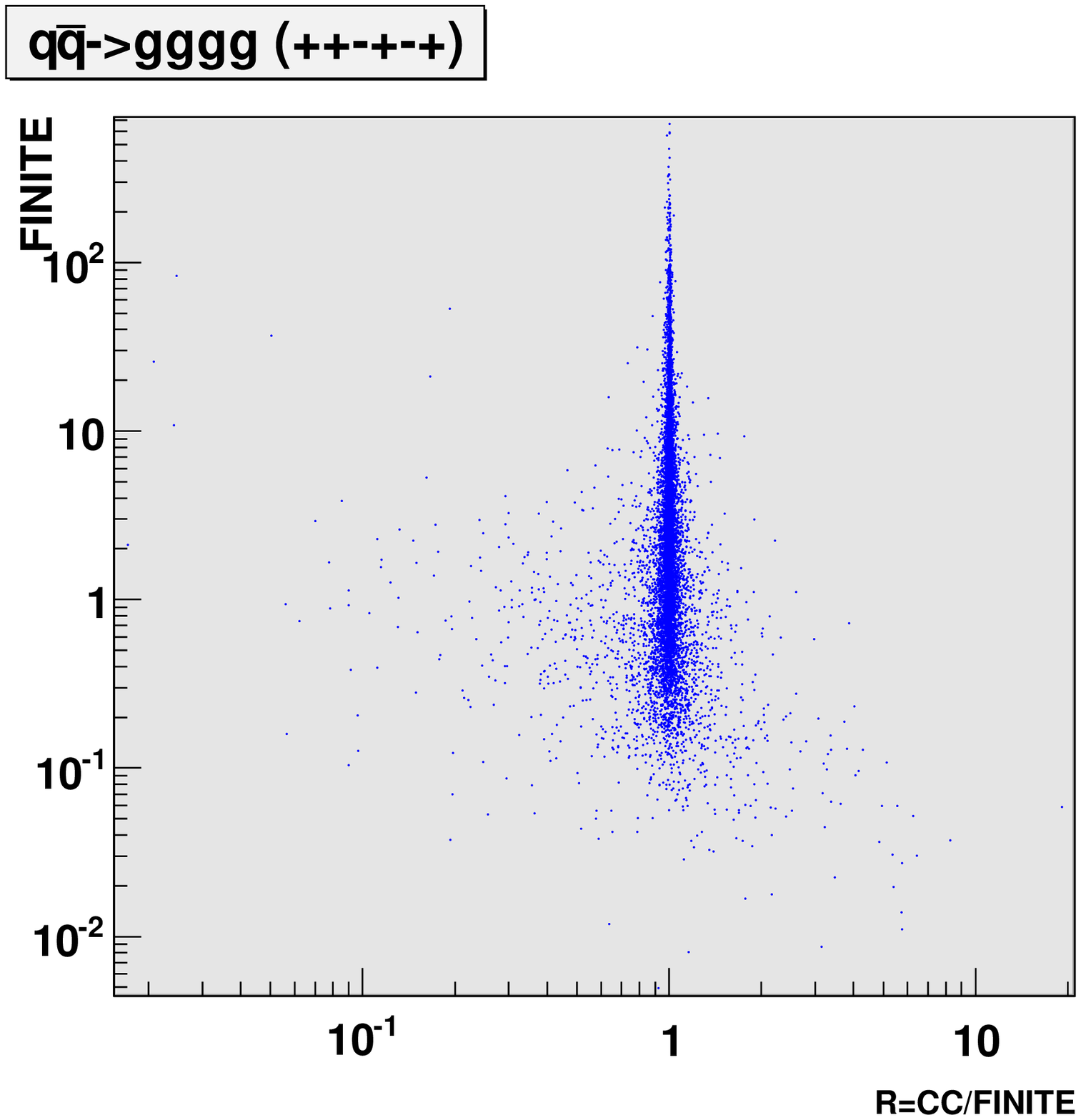,width=0.47\textwidth}
{
Scatterplot of the absolute value of the total finite result for the primitive amplitude, $|FIN|$ versus the ratio $|CC| /  |FIN|$ for the $++-+--$ helicity configuration of the $q\bar{q}\rightarrow 4g$ leading color primitive. See section \ref{Kinematical_cuts} for details about the generation of the phase space pointset.
\label{cc_over_finite_3}
}
{
Scatterplot of the absolute value of the total finite result for the primitive amplitude, $|FIN|$ versus the ratio $|CC| /  |FIN|$ for the $++-+-+$ helicity configuration of the $q\bar{q}\rightarrow 4g$ leading color primitive. See section \ref{Kinematical_cuts} for details about the generation of the phase space pointset.
\label{cc_over_finite_4}
}


\DOUBLEFIGURE
{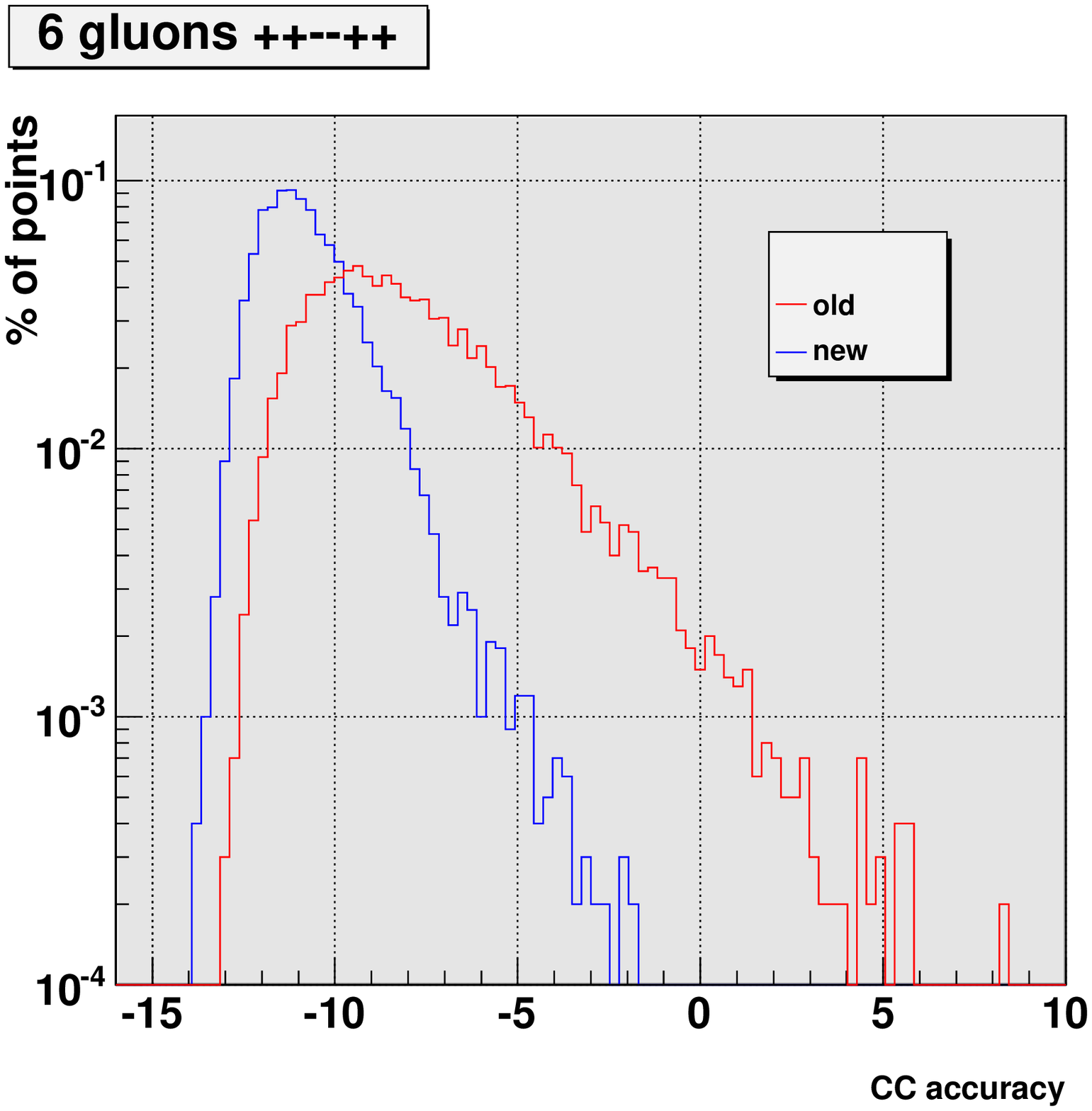,width=0.47\textwidth}
{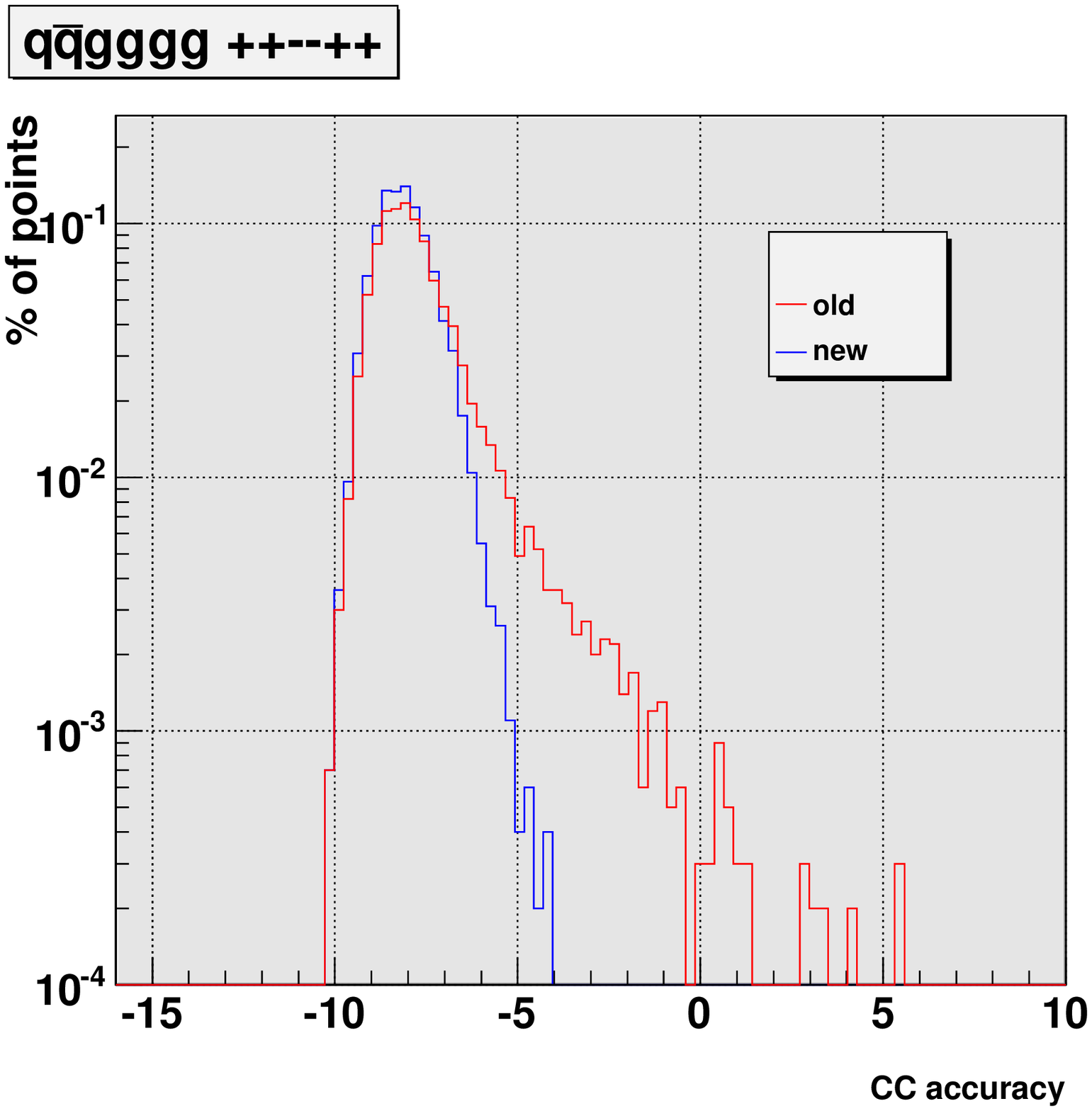,width=0.47\textwidth}
{
CC in the canonical and the new approach pole in $N=6$ gluons.See section \ref{Kinematical_cuts} for details about the generation of the phase space pointset.
\label{cc_PID_1}
}
{
CC in the canonical and the new approach pole in $q\bar{q}\rightarrow 4g$.See section \ref{Kinematical_cuts} for details about the generation of the phase space pointset.
\label{cc_PID_4}
}


\DOUBLEFIGURE
{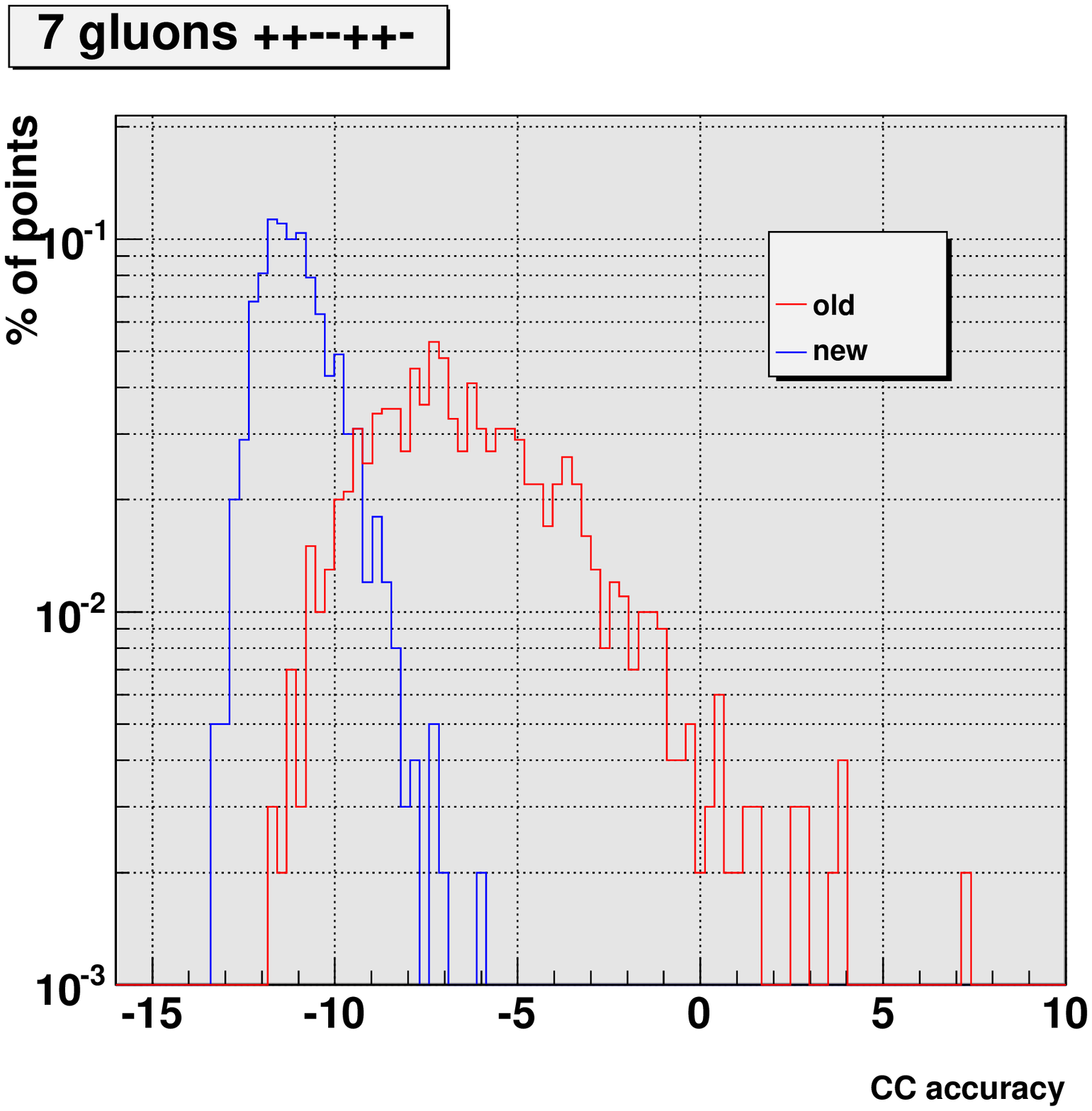,width=0.47\textwidth}
{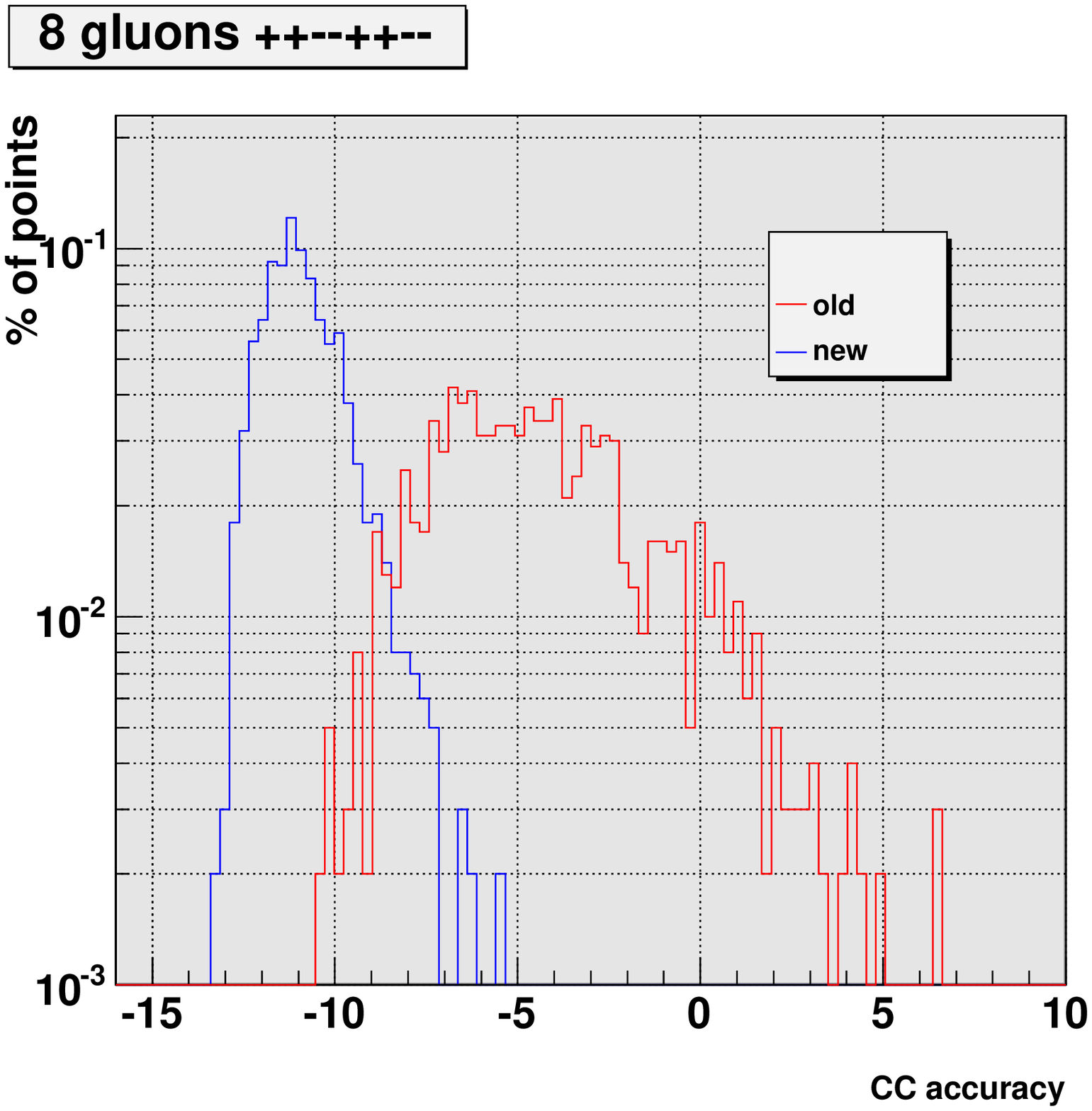,width=0.47\textwidth}
{
CC in the canonical and the new approach pole in $N=7$ gluons. See section \ref{Kinematical_cuts} for details about the generation of the phase space pointset.
\label{cc_PID_7}
}
{
CC in the canonical and the new approach pole in $N=8$ gluons. See section \ref{Kinematical_cuts} for details about the generation of the phase space pointset.
\label{cc_PID_12}
}


The importance of reconstructing the CC part of the amplitude with a numerically accurate method is amplified by the fact that, in the majority of phase space points, the CC part gives the larger contribution to the finite part of the amplitude.  In fig.\ref{cc_over_finite_1}-\ref{cc_over_finite_2} we show scatterplots of the total $finite$ result versus the ratio of  $CC/finite$ for the $gg\rightarrow 4g$ primitive for two different helicity configurations, and in fig.\ref{cc_over_finite_3}-\ref{cc_over_finite_4} for the $q\bar{q}\rightarrow 4g$ leading color primitive. The percentage of points where the ratio deviates from 1 by, say, $10\% $ depends on the particular helicity configuration, but it is clear that for the majority of points  (from $75\%$ for pure gluonic primitives to $95\%$ for primitives that include fermions) the CC part is the larger contribution to the finite result for the primitive amplitude.

\section{The rational part}
\label{Kinematical_cuts}
When the loop momentum $l_s$ is five-dimensional, the coefficients $d_{i,s}^{b},c_{i,s}^{b},b_{i,s}^{b}$ are non-zero, and moreover depend on the numerically dangerous pentagon coefficients linearly. But knowledge of their explicit functional form allows us to reconstruct them in a way that partially  avoids numerical cancelations.

To this effect one notes that 
\be
\bar{d}_{I_4,s}^{b}=-\sum_{k \in J_5 / I_4} \frac{\bar{e}_{k}}{D_{k}(l_s)} = -\sum_{k \in J_5 / I_4} \frac{\bar{e}_{k}}{(l_s+q_k)^2}
\ee
and, since for the box we have $l_s^{\mu}=V^{\mu}+a_1 n^{\mu}$ (for which $l_s^2=0$),
\be
(l_s+q_k)^2=q^2+2 l_s\cdot q_k = q^2+ 2 V\cdot q_k + a_1 n \cdot q_k \equiv x + a_1 y 
\ee
So
\be
\bar{d}_{I_4,s}^{b}=-\sum_{k \in J_5 / I_4} \frac{ \bar{e}_{k} } { x_{I_4}+a_1(l_s) y_{I_4} } 
\ee
from which
\be
d^b_{I_4,0}=\sum_{k \in J_5 / I_4}  \bar{e}_{k} \frac{x_{I_4}}{a_{I_4}^2y_{I_4}^2-x_{I_4}^2}
\ee
\be
d^b_{I_4,1}=\sum_{k \in J_5 / I_4}  \bar{e}_{k} \frac{-y_{I_4}}{a_{I_4}^2y_{I_4}^2-x_{I_4}^2}
\ee
with $a_{I_4}\equiv \sqrt{-V_{I_4}^2}$.
Similar expressions can be found  for $d_{2,3,4}$.  
The box counter term functions $d^{b}_{J_4}(l)$ (needed for the triple-cuts) can then be reconstructed for any $l$, with the pentagon coefficient  pulled out as a common factor. This prevents cancelations of large terms while solving the system.

\DOUBLEFIGURE
{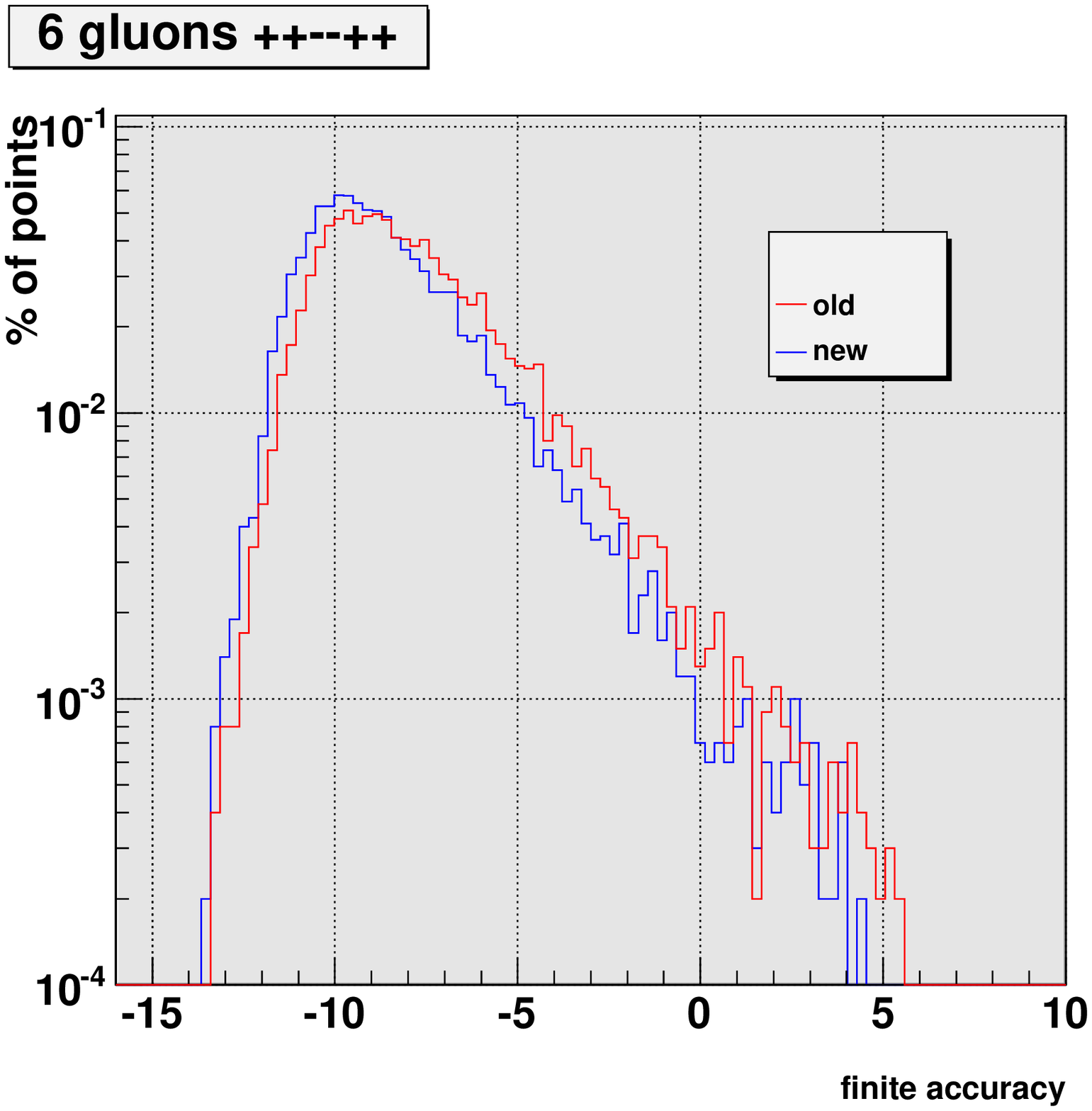,width=0.47\textwidth}
{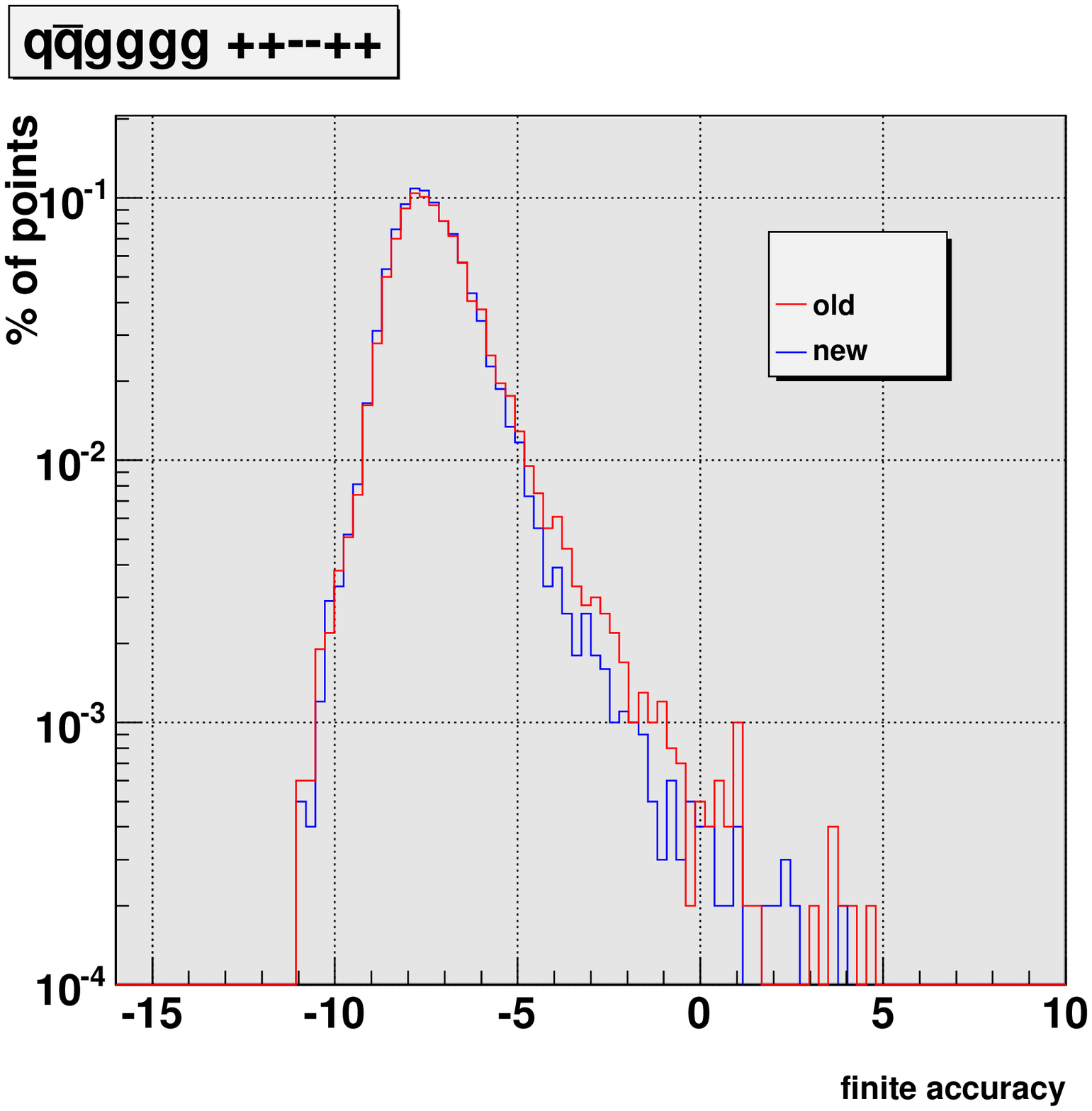,width=0.47\textwidth}
{
The total finite part with the canonical and the new approach in $N=6$ gluons. See section \ref{Kinematical_cuts} for details about the generation of the phase space pointset.
\label{total_finite_PID_1}
}
{
The total finite part with the canonical and the new approach in $q\bar{q}\rightarrow 4g$ gluons. See section \ref{Kinematical_cuts} for details about the generation of the phase space pointset.
\label{total_finite_PID_4}
}


However, in the majority of the problematic points, the rational part is inaccurately reconstructed due to the inaccurate evaluation of the  the pentuple-cut residue.  The above treatment therefore, only partially  improves the numerical accuracy of the rational part, as shown in figs.\ref{total_finite_PID_1}-\ref{total_finite_PID_4} where the accuracy of the rational part relative to the rational part calculated with quadruple precision is shown with and without the pentagon isolation approach described above. We expect that evaluating the pentuple cuts only, in higher precision, would remedy most of the problematic points left.

Phase space points are selected uniformly by an implementation of the RAMBO\cite{Kleiss:1985gy} algorithm, with  cuts $E_T>10^{-2}\sqrt{s}$,$|n|<4$,$\Delta R=\sqrt{\Delta n^2+\Delta \phi^2}>0.7$. The definition of accuracy that we adopt is the standard one: 
\be
\epsilon=\log_{10}{\frac{\left | x_N-x_P \right |}{\left | x_P\right |}}
\ee
where $x$ can be the coefficient of the double or single pole in the $\epsilon$-expansion of the primitive, or the CC part or the total finite part, $x_N$ is the numerical value of $x$ and $x_P$ is the predicted value, either from the analytic expression for the poles \cite{Catani:1998bh,Giele:1991vf,Kunszt:1994mc} or from numerical evaluation in quadruple (64 digits) precision, for the finite part.

\section{Computational efficiency improvement}
Thanks to the improved evaluation of the cut constructible  part of the amplitude, the number of points for which one needs to resort to quadruple precision arithmetics is reduced drastically. In practice one can avoid quadruple precision in $20-35\%$ of the previously problematic points. Since evoking quadruple precision typically involves a cpu penalty factor of the order of 100, the improvement can be significant in terms of total cpu time, particularly in the case of pure gluonic amplitudes or amplitudes with one fermion pair. For example, in the case of the $q\bar{q}\rightarrow gggg$ used as an example above, with the tight cuts we have adopted in this work, with the canonical approach one has $3.6\%$ of the phase space points with relative accuracy of less than  
$10^{-4}$. The new approach brings the percentage down to $2.3\%$. This translates into a $30\%$ reduction of the necessary cpu time needed to evaluate all points with relative accuracy better than $10^{-4}$.

\section{Conclusions}
We have suggested a new method to disentangle the contribution of pentuple-cuts from the evaluation of the Cut Constructible part of one loop amplitudes within the framework of D-dimensional unitarity without increasing the computational complexity of the algorithm. We have showed that in all cases studied the Cut Constructible part is now evaluated with high precision. Knowledge about the explicit functional form of the contributions to the rational part that originate from pentuple-cuts allows for a small improvement in its evaluation. As a result, the percentage of phase-space points for which one needs to resort to higher precision arithmetics (which is computationaly expensive)  in order to reconstruct the virtual amplitude correctly is drastically reduced. This leads to a significant improvement in the efficiency of the D-dimensional unitarity approach without the need for restricting the phase space available for final state jets or vetoing of problematic phase space points. 

\acknowledgments
I would like to thank Zoltan Kunszt and Babis Anastasiou for stimulating discussions on unitarity techniques and related issues.

\bibliographystyle{JHEP}
\bibliography{PreciseCC}

\end{document}